\documentclass{nature}
\usepackage{pifont}
\usepackage{bbding}
\usepackage{graphicx}
\usepackage{hyperref}
\usepackage{bbm}
\usepackage{mathrsfs}
\usepackage{graphicx}
\usepackage{dcolumn}
\usepackage{bm}
\usepackage{amsmath}
\usepackage{amsfonts}
\usepackage{caption}
\usepackage{color}

\title{Predicting Many Crystal Properties via an Adaptive Transformer-based Framework}

\author{Haosheng Xu$^{1,2}\dag$ Dongheng Qian$^{1,2}\dag$ Jing Wang$^{1,2,3,4\ast}$}

\begin{document}

\maketitle

\begin{affiliations}
\item State Key Laboratory of Surface Physics and Department of Physics, Fudan University, Shanghai 200433, China
\item Shanghai Research Center for Quantum Sciences, Shanghai 201315, China
\item Institute for Nanoelectronic Devices and Quantum Computing, Zhangjiang Fudan International Innovation Center, Fudan University, Shanghai 200433, China
\item Hefei National Laboratory, Hefei 230088, China\\
\textnormal{\textsuperscript{\dag} These authors contributed equally to this work. \\
\textsuperscript{*} Corresponding author: wjingphys@fudan.edu.cn}
\end{affiliations}

\begin{abstract}
{\bf Machine learning has revolutionized many fields, including materials science. However, predicting properties of crystalline materials using machine learning faces challenges in input encoding, output versatility, and interpretability. We introduce CrystalBERT, an adaptable transformer-based framework integrating space group, elemental, and unit cell information. This novel structure can seamlessly combine diverse features and accurately predict various physical properties, including topological properties, superconducting transition temperatures, dielectric constants, and more. CrystalBERT provides insightful interpretations of features influencing target properties. Our results indicate that space group and elemental information are crucial for predicting topological and superconducting properties, underscoring their intricate nature. By incorporating these features, we achieve 91\% accuracy in topological classification, surpassing prior studies and identifying previously misclassified materials. This research demonstrates that integrating diverse material information enhances the prediction of complex material properties, paving the way for more accurate and interpretable machine learning models in materials science.}
\end{abstract}

\section*{Introduction}
Machine learning (ML) has emerged as a prominent research field in computer science, extending its applications to  various domains within physics~\cite{carleo2019}. A significant application is in uncovering patterns and predicting properties from large datasets, which might be previously unknown or difficult for humans to calculate~\cite{snyder2012,venderley2018,zhang2017}. In material science and condensed matter physics, ML models have been developed to predict a wide range of material properties without the need for specific, arduous calculations~\cite{agrawal2016,choudhary2022,vasudevan2019,schmidt2019,schutt2018,deringer2018, grisafi2018,magar2022a,magar2022b,cao2023, jinnouchi2019,chen2019,ock2023,kazeev2023,keqiang2022,choudhary2018,ward2018}.

When designing ML models, several challenges typically arise. The first challenge involves input encoding, which concerns how to represent a specific material as vectors. Previous strategies can be broadly categorized into three approaches: manual feature selection~\cite{claussen2020,tao2021}, text description which describes a material using a human-readable sentence for input into a large language model~\cite{huang2023a,choudhary2024,gupta2022,rubungo2023,olivetti2020}, and physically motivated methods such as crystal graphs~\cite{xie2018,choudhary2021,karamad2020}. A pertinent question is whether a general method exists to integrate all these types of features into a single model. Recent work has begun to explore this issue by applying multimodal learning~\cite{gong2023a,gong2023b,huang2024}. The second challenge is output versatility. Most models excel at certain tasks but perform poorly on others. Given the goal of predicting various properties of a material, it is worth exploring whether a general model can perform well across a wide range of tasks. Data utilization presents the third significant challenge. For specific tasks, the number of labeled materials might be relatively small compared to the total number of known materials. A model that effectively utilizes vast amounts of unlabeled data would be desirable and potentially more powerful, making pretrained model a promising solution. Interpretability is the final aspect. Beyond knowing what the model predicts, it is essential to understand why it makes certain predictions. While deep learning models are often viewed as black boxes, making it challenging to derive a concrete formula relating input to output, there is still a need to identify the most relevant features that influence a specific property. 

In this Letter, we introduce CrystalBERT (XBERT), a novel ML model designed to potentially address  all the aforementioned challenges. The model structure is illustrated in Fig. 1(a). We utilize the Bidirectional Encoder Representation from Transformers (BERT) architecture~\cite{devlin-etal-2019-bert} for pretraining our model, treating the prediction of various physically important properties as different downstream tasks. During pretraining, we leverage large amounts of crystalline materials to develop a sophisticated representation of materials, subsequently using smaller labeled datasets for specific downstream tasks. To our knowledge, XBERT presents a novel model structure by being the first to integrate graph neural network (GNN) outputs with global properties using a transformer architecture. We treat space group information and other global properties as tokens, and distinctively incorporate the output of the Crystal Graph Convolutional Neural Network (CGCNN)~\cite{xie2018} or the Atomistic Line Graph Neurla Network (ALIGNN)~\cite{choudhary2021} as an additional input token. This method allows for a more comprehensive and detailed encoding of materials, addressing the key challenges of input encoding, output versatility, data utilization, and interpretability in ML models for material science.

We deliberately select a diverse set of properties to predict, including topological properties, superconducting transition temperatures, dielectric constants, and more. Our model’s performance is better than or comparable to the underlying GNN results across all these tasks. Remarkably, we achieve around 91\% accuracy rate for topological property prediction, outperforming all previous results~\cite{claussen2020,schleder2021,ma2023,Andrejevic2022,Liu2021,Cao2020,xu2024,rasul2023}, and even identifying some previously misclassified topological materials. We further provide insightful physical interpretations regarding the features that most significantly influence the target properties. These findings pave the way for developing more powerful ML models to predict material properties and are of significant physical importance.

\section*{Results}
\paragraph*{XBERT model} To construct the input to our XBERT model, each crystalline material is described by eight tokens. The first five tokens, inspired by~\cite{huang2023a}, encode space group information, including space group symbol, point group, crystal system, Laue class and symmetry. For example, if the space group of a material is 225, then the five tokens are $[F\frac{4}{m}\Bar{3}\frac{2}{m}],[m\Bar{3}m],[\text{cubic}],[m\Bar{3}m]$ and $[\text{Centrosymmetric}]$. As is commonly done in natural language processing, we counted all the `words' that appeared in the tokens derived from the 230 space groups and removed the duplicates to get a dictionary of length 280. Then each 'word' was assigned a number (from 0 to 279), which can be easily embedded into a 256-dimensional vector. The sixth token represents global properties; in this work, we include only the total electron number in a unit cell($N_{e}$), though it is straightforward to add additional properties such as the volume of the unit cell and the number of atoms in a unit cell. $N_{e}$ is represented by the digits in its binary representation. The largest $N_{e}$ in our dataset is 11,448 which can be represented by a 14-dimensional vector $[1,0,1,1,0,0,1,0,1,1,1,0,0,0]$. Thus, for each material, $N_{e}$ is described by a 14-dimensional vector. Leading zeros are added to the vector if its binary representation is less than 14 bits. The seventh token encodes the element information directly in the form of $\sum_{\mu}f_{\mu}\vec{v_{\mu}}$, where $f_{\mu}$ and $\vec{v_{\mu}}$ represent the proportion of element $\mu$ in a material and element vector of element $\mu$ taken from \cite{xie2018}, respectively. The eighth token represents the output of a GNN model. We primarily consider using the CGCNN model~\cite{xie2018}. Initially, each atom in the unit cell is assigned a vector $\mathbf{v}_{i}^{0}$ , which is solely determined by the element type. The edge between every two atoms is described by a vector $\mathbf{u}_{ij}$, encoding the distance between the two atoms. Subsequently, through the message passing mechanism, information from the surrounding atoms will be transmitted to the target atom, resulting in the updating of the atom's vector:
\begin{equation}\label{equ1}
\mathbf{v}^{t+1}_{i}=\mathbf{v}^{t}_{i}+\sum_{j,k}\sigma(\mathbf{z}^{t}_{(i,j)_{k}}\mathbf{W}^{t}_{f}+\mathbf{b}^{t}_{f})\odot g(\mathbf{z}^{t}_{(i,j)_{k}}\mathbf{W}^{t}_{s}+\mathbf{b}^{t}_{s})
\end{equation}
\begin{equation}\label{equ2}
\mathbf{z}^{t}_{(i,j)k}=\mathbf{v}^{t}_{i}\oplus \mathbf{v}^{t}_{j}\oplus \mathbf{u}_{(i,j)_{k}}
\end{equation}
where $\odot$ and $\oplus$ denote element-wise multiplication and concatenation of feature vectors, respectively. After $L$ rounds of message passing, the vector on each atom is updated to be $\mathbf{v}^{L}_{i}$. We then take the average of the vectors of all atoms within the unit cell to obtain a representation of the entire unit cell:
\begin{equation}\label{equ3}
\mathbf{v} = \frac{1}{N}\sum_{i=1}^{N}\mathbf{v}_{i}^{L}.
\end{equation}
The parameters in CGCNN are also updated during training. Every token is embedded into a vector of length 256 before being fed into the transformer encoder. This method effectively combines the three aforementioned encoding approaches in a concrete and flexible manner. To demonstrate the flexibility of XBERT and to achieve superior performance, we train a model in which CGCNN is replaced by a more advanced graph neural network, ALIGNN~\cite{choudhary2021}. We anticipate that employing a more powerful GNN model will further enhance the performance of our XBERT model, as discussed later.

The architecture of our XBERT model is based on the encoder side of the transformer~\cite{vaswani2017}, with the core being the self-attention layers as shown in Fig. 1(a) and Fig. 1(b). Three different embeddings (queries ($Q$), keys ($K$) and values ($V$)) are obtained from the input tokens. The attention score is then calculated using the formula:
\begin{equation}\label{equ0}
    \text{Attention}\left(Q,K,V\right) = \text{softmax}\left(\frac{QK^{T}}{\sqrt{d_{k}}}\right)V,
\end{equation}
where $d_{k}$ is the dimension of keys $K$. XBERT utilizes 5 attention layers with 4 heads. The output vectors $\{\vec{x_{i}}\}$ from the last layer are concatenated into a single long vector, referred to as the crystal feature vector. These crystal feature vectors serve as the input to the subsequent fully connected layer, whose outputs are then used to predict various properties depending on the specific downstream task.

During pretraining, we use approximately 150,000 crystalline materials from the Materials Project~\cite{jain2013}. The pretraining task involves predicting the lattice parameters of the unit cell, which aids our model in capturing the complex interplay between structural and elemental information. In Fig. 1(c), we apply t-distributed stochastic neighbor embedding (t-SNE) dimension reduction to the crystal feature vectors output by the pretrained model and plot them in a 2D space~\cite{laurens2008}. Different colors represent different crystal systems, and materials belonging to the same crystal system cluster together, demonstrating that the pretraining task has been effectively accomplished. For fine-tuning, we utilize the weights from the pretrained model as the initial weights. Detailed training methods can be found in Methods.

\begin{table*}[t]
	\caption{\textbf{Summary of the prediction performance for ten different models on eleven different properties~\cite{dunn2020}.} The values represent accuracy for topological classification and mean absolute value (MAE) for other tasks. The MAE values are obtained from 5-fold cross-validation and rounded to three significant figures. Different models are defined as: XBERT$_\text{FC}$ refers to the full model (with CGCNN); I, II, III, IV excludes the transformer, macroscopic properties, element information, CGCNN, respectively; V is XBERT$_\text{FC}$ without the softmax layer; and VI is II without the softmax layer. XBERT$_\text{FA}$ refers to the full model (with ALIGNN). TopoA, TopoB, 3DSC, Perovskites, Dielectric, GVRH, JDFT2D, KVRH, Band Gap, Phonons, and E-Form are datasets representing topological materials in Ref.~\cite{Po2017, Tang2019_1, Tang2019_2}, in Ref.~\cite{Bradlyn2017, Vergniory2019, Maia2022}, superconducting transition temperature~\cite{Sommer2023}, formation energy of perovskites~\cite{castelli2012}, refractive index~\cite{petousis2017}, shear modulus~\cite{dejong2015, ward2018}, exfoliation energy~\cite{choudhary2017}, bulk modulus~\cite{dejong2015}, band gap~\cite{jain2013}, last phdos peak~\cite{petretto2018}, and formation energy~\cite{jain2013},  respectively. The best results achieved in XBERT$_\text{FC}$-based models are shown in blod. }
	\begin{center}\label{results}
        {\footnotesize
		\renewcommand{\arraystretch}{0.8}
		\begin{tabular*}{6in}
			{@{\extracolsep{\fill}}ccccccc}
			\hline
			\hline
			Model &TopoA & TopoB & 3DSC & Perovskites & Dielectric & GVRH\\
			\hline
			CGCNN & 85.49\% & 86.28\% & 3.17(46) & 0.048(3) & 0.541(66) & 0.086(2)   \\  
			XBERT$_\text{FC}$ & \textbf{89.76\%} & 90.91\% & \textbf{2.69(25)} & 0.157(32) & \textbf{0.378(66)} & 0.094(4) \\
            I & 86.76\% & 89.10\% & 3.06(33) & 0.291(11) & 0.400(41) & 0.090(1) \\
			II & 88.14\% & 87.85\% & 2.74(36) & 0.143(14) & 0.388(24) & 0.091(2)  \\ 
			III & 89.74\% & 90.86\% & 2.76(35) & 0.169(36) & 0.397(51) & 0.093(3)  \\ 
			IV & 88.83\% & 89.84\% & 2.90(34) & 0.457(12) & 0.435(103) & 0.121(4)  \\ 
			V & 89.56\% & \textbf{91.19\%} & 2.85(49) & 0.079(3) & 0.488(57) & 0.088(2)  \\ 
			VI & 87.52\% & 88.17\% & 2.74(31) & \textbf{0.076(3)} & 0.455(54) & \textbf{0.085(3)}  \\
            \hline
            ALIGNN & 87.91\% & 88.36\% & 2.67(42) & 0.029(1) & 0.385(32) & 0.072(1)  \\
            XBERT$_\text{FA}$ & 89.90\% & 91.33\% & 2.59(44) & 0.033(1) & 0.335(49) & 0.071(1) \\
			\hline
			\hline
        \end{tabular*}}
    \end{center}
    \label{table1}
\end{table*}

\newpage
    
\begin{table*}[t]
    \caption*{Table 1 Continued.}
    \begin{center}\label{results_2}
        {\footnotesize
    	\renewcommand{\arraystretch}{0.8}
    	\begin{tabular*}{6in}
    		{@{\extracolsep{\fill}}cccccc}
            \hline
			\hline
            Model & JDFT2D & KVRH & Band Gap & Phonons & E-Form   \\
            \hline
            CGCNN & 46.7(6.1) & 0.070(4) & 0.263(8) & 56.1(4.8) & 0.042(3)   \\
            XBERT$_\text{FC}$ & 51.7(5.1) & 0.078(7) & 0.244(8) & 93.0(8.6) & 0.047(2)   \\
            I & 64.0(3.7) & 0.073(3) & 0.284(5) & 157.0(7.8) & \textbf{0.044(3)}   \\
            II & 52.3(7.1) & 0.077(1) & 0.273(10) & 99.2(15.0) & 0.049(4)   \\
            III & 53.7(3.6) & 0.079(4) & \textbf{0.235(6)} & 109.8(20.0) & 0.048(3)   \\
            IV & 54.8(4.2) & 0.096(6) & 0.321(6) & 98.6(10.1) & 0.105(4)   \\
            V & 49.1(3.3) & 0.071(3) & 0.262(21) & 78.8(10.0) & 0.045(4)   \\
            VI & \textbf{48.2(8.4)} & \textbf{0.069(3)} & 0.258(10) & \textbf{61.7(3.9)} & 0.046(4)   \\
            \hline
            ALIGNN & 42.9(9.9) & 0.058(1) & 0.191(4) & 40.5(7.4) & 0.021(1) \\
            XBERT$_\text{FA}$ & 40.5(10.9) & 0.059(3) & 0.211(11) & 51.1(10.1) & 0.034(2)  \\
            \hline
			\hline
		\end{tabular*}}
	\end{center}
\end{table*}

\paragraph*{Performance and analysis} To evaluate XBERT's performance and demonstrate its generality, we selected ten different material properties that encompass various aspects of crystalline materials as prediction targets, as detailed in Methods. We assessed the model's performance by removing different input features to understand their influence on specific properties. Additionally, we explored various model structures, such as replacing the transformer with a simple fully connected layer, and examining whether normalizing each output token vector using a softmax layer affects the results. The final outcomes are presented in Table. 1. 

By comparing the performance of models with and without the transformer, it is evident that the transformer structure is essential for achieving good performance. The XBERT-based model outperforms CGCNN in seven tasks, and the model with all input features outperforms CGCNN in five tasks. As shown in Supplementary Table S3, the token corresponding to CGCNN has the highest sum of absolute values before normalization. Thus, removing the normalization layer amplifies the impact of CGCNN and improves model performance for tasks that heavily depend on unit cell information. The inferior performance of the full model in some tasks can be attributed to less important features detracting from overall performance of our model, despite its ability to prioritize relevant features. Therefore, better performance is expected with fewer input features. For predicting bulk modulus, removing the electron number feature yields better results, which is intuitive since electron number is irrelevant to modulus. Similarly, removing the elemental feature achieves the best performance for the electron gap task. For tasks related to phonons, the properties are primarily dependent on unit cell information, resulting in our model's performance being slightly inferior to CGCNN. 

Notably, in predicting topological properties and superconducting properties, XBERT shows clear advantages over CGCNN. The substantial decrease in performance upon removing either space group, elemental, or unit cell information indicates that these properties result from the intricate interplay between structural and elemental information. To further enhance performance, we also trained an XBERT model by replacing CGCNN with ALIGNN, expecting improved outcomes. As anticipated, the performance significantly exceeded that achieved with CGCNN across all tasks, and our model outperformed ALIGNN on six tasks. These findings suggest that the XBERT-based model would likely benefit further from more advanced GNN architectures, and it is expected to maintain its advantages in predicting complex properties over GNN-based models.

To better interpret the model, we can investigate its workings through two steps. First, we aim to understand what information the output token encodes.  We simplify the fine-tuning layer by mapping each token vector $\vec{x_i}$ to a single value $s_i$ and consider the final output as $y=\sum_i{s_i}$. Taking the topological property task as an example, we focus on the value $s_{sg}$ corresponding to the space group token and aim to understand its physical meaning. By averaging this value over materials in the same space group, we obtain the mean feature value $\overline{s_{sg}}$ for each space group. As illustrated in Fig. 2(a), this value actually reflects the proportion of topological materials within each space group. In \cite{claussen2020}, it was found that this proportion is a better encoding of space group information and improves performance. Importantly, this feature is automatically learned by our model rather than being manually instructed by human intelligence. Furthermore, we found that higher group numbers usually have a higher value of topological contribution, which may be due to the fact that more symmetry provides a greater ability to protect certain kinds of topology. To visualize this relationship, we present in Fig.3 a color-coded representation of the $\overline{s_{sg}}$ values, where darker colors represent higher $\Bar{s}_{sg}$ values. Space groups that appear fewer than 20 times in the materials were excluded from the analysis. From the figure, we observe a general trend where, within the same crystal system, the color tends to darken as the space group number increases.

Next, we aim to determine the influence of each feature on specific physical properties. As demonstrated in the Supplementary Materials, while the attention layers in the transformer encoder allow each output token to access information from all other tokens during both pretraining and fine-tuning, the output tokens are predominantly influenced by their corresponding input tokens due to the residual connections~\cite{kaiming2016}. Consequently, each output token can be regarded as primarily encapsulating the feature of its respective input token. Our model has only one linear layer connecting the output vector of the transformer encoder to the final output, represented by $y = \sum_{ij}w_{ij}x_{ij} + b_{ij}$. Here, $i$ represents different tokens, $j$ labels different components in the token vector, and $x_{ij}$ is the crystal feature vector. The value $\sum_j{\left| w_{ij} \right|}$ reflects the influence that feature $i$ has on the property, as shown in Fig. 2(b-d). For formation energy, where unit cell information is commonly believed to be more important, the highest weight is on the last token. Remarkably, for topological properties and superconducting transition temperature, the weights are more concentrated on the space group and element tokens. This indicates that these features are more significant and emphasizes the intricate nature of these properties, which cannot be fully captured by CGCNN.

\paragraph*{Advantages} Since we have already demonstrated the generality and interpretability of XBERT, we now address two additional advantages of our model: efficiency and accuracy. First, XBERT's efficiency is a notable advantage, which is commonly observed in pretrained models~\cite{devlin-etal-2019-bert}. We compare the train loss reduction for different downstream tasks between XBERT and CGCNN. Across all tasks, XBERT demonstrates a faster reduction in loss, as shown in Fig. 4, highlighting its superior efficiency.

Second, XBERT achieves exceptional accuracy, particularly in predicting topological properties, with $89\%$ accuracy for TopoA~\cite{Po2017,Tang2019_1,Tang2019_2} and $91\%$ for TopoB~\cite{Bradlyn2017,Vergniory2019,Maia2022}. Additionally, Table.2 show the precision, recall and F1 score for the TopoA and TopoB tasks, with their definition explained in Materials and Methods. We can easily find that XBERT performs better than CGCNN in all metrics.
For the TopoB task, we used Model V, which performed best in the binary classification, to conduct a three-class classification of topological insulators, topological semimetals, and trivial materials. We compared this with the results of directly using CGCNN for the three-class classification. As shown in Table.3, XBERT achieved higher accuracy and F1 scores for each category compared to CGCNN. However, as indicated by the F1 scores, our model's ability to distinguish topological insulators was lower than its performance for topological semimetals and trivial materials, which is consistent with the findings in Ref.~\cite{claussen2020}. It is worth emphasizing that Ref.~\cite{rasul2023} tested various other graph neural networks, including GATGNN, DeeperGATGNN, and MEGNet, as well as their proposed models, for this three-class classification task. None of these models achieved an accuracy exceeding 80\%. This benchmark indicates that combining additional information with the graph neural network through the Transformer encoder layers can significantly improve the accuracy of the three-class classification.

\begin{table*}[htbp]
	\caption{\textbf{Accuracy, precision, recall and F1 score for the TopoA ant TopoB task.} The following results are the average values from cross-validation. The best results are shown in bold.}
	\begin{center}
		\renewcommand{\arraystretch}{0.8}
		\begin{tabular*}{5.5in}
			{@{\extracolsep{\fill}}c|ccc|ccc}
			\hline
			\hline
                  &     &  TopoA  &    &    &  TopoB &   \\
			Model & Precision & Recall & F1 score & Precision & Recall & F1 score\\
			\hline
			CGCNN & 85.03\% & 85.20\% & 84.87\% & 86.91\% & 87.08\% & 86.80\%\\  
			$\text{XBERT}_{\text{FC}}$ & \textbf{90.42\%} & 88.11\% & 89.10\% & 91.03\% & \textbf{91.72\%} & 91.25\%\\
            I & 86.69\% & 86.09\% & 86.18\% & 88.93\% & 90.54\% & 89.59\%\\
			II & 88.15\% & 87.38\% & 87.60\% & 88.20\% & 88.84\% & 88.35\%\\ 
			III & 90.21\% & \textbf{88.48\%} & \textbf{89.18\%} & 91.10\% & 91.52\% & 91.20\%\\ 
			IV & 89.34\% & 87.42\% & 88.18\% & 89.93\% & 90.89\% & 90.27\%\\ 
			V & 90.20\% & 87.97\% & 88.89\% & \textbf{91.83\%} & 91.35\% & \textbf{91.46\%} \\ 
			VI & 87.27\% & 87.11\% & 87.01\% & 88.61\% & 88.27\% & 88.27\%\\ 
			\hline
			\hline
		\end{tabular*}
	\end{center}
 \label{table2}
\end{table*}

\begin{table*}[htbp]
	\caption{\textbf{Performance of CGCNN and model V as measured by the accuracy and F1 score.} The F1 score for three-class classification is calculated using the harmonic mean of precison and recall for each class. }
	\begin{center}
		\renewcommand{\arraystretch}{0.8}
		\begin{tabular*}{5.5in}
			{@{\extracolsep{\fill}}ccccc}
			\hline
			\hline
			Model &Accuracy & F1 Triv. & F1 TI & F1 TSM\\
			\hline
			CGCNN & 74.03\% & 85.08\% & 42.60\% & 71.34\%\\  
			V & 85.71\% & 90.60\% & 63.51\% & 88.06\% \\ 
			\hline
			\hline
		\end{tabular*}
	\end{center}
 \label{table3}
\end{table*}

This high level of accuracy allows us to identify previously undiscovered or misclassified topological materials. Since symmetry indicator already serves as a quick and reliable check for topological material~\cite{kruthoff2017,Zhang2019}, our goal is to identify non-symmetry-diagnosable and previously undiscovered topological materials. Here we focus on non-magnetic topological insulators. To achieve this, we conducted screening on two different datasets. 
Firstly, we employ XBERT to scan through 1,433 materials that are undiagnosable by symmetry indicators, as outlined in~\cite{ma2023}, and identify 206 materials labeled as topological. We subsequently exclude those that are already known to be topological and those that are metals. As a result, we find that Ag$_{2}$HgSe$_{4}$Sn is a promising topological insulator candidate which has a large band gap of about $16.2$~meV. 
Fig. 5(a) illustrates the crystal structure. Fig. 5(b) and 5(c) reveal that spin-orbit coupling opens a gap at $\Gamma$ point, leading to nontrivial topology, which is confirmed by Dirac surface state shown in Fig. 5(d). Interestingly, this material has already been experimentally fabricated~\cite{parasyuk2002} and was misclassified by the ML method in~\cite{xu2024,ma2023} since its topogivity is around $-7$, which should be regarded as quite trivial. 
Additionally, we observed that in the TopoB dataset, topological materials lacking inversion symmetry may not be identifiable by symmetry indicator, potentially leading to their misclassification as trivial. To address this, we conducted a screening of materials classified as trivial but lacking inversion symmetry, in search of potential topological insulators. This approach resulted in the identification of 10 additional topological insulators, which are listed in Table 4. Their band structures and Wilson loops, along with further training details, are provided in the Supplementary Materials.

\begin{table*}[htbp]
	\caption{\textbf{The ten newly discovered topological insulators.} These materials were not identified as topological in database established in Ref.~\cite{Bradlyn2017,Vergniory2019,Maia2022}, but discovered by our XBERT.}
	\begin{center}
		\renewcommand{\arraystretch}{0.8}
		\begin{tabular*}{5.5in}
			{@{\extracolsep{\fill}}cccccc}
			\hline
			\hline
			ICSD &Formula & Space group & Type & Gap (meV)\\
			\hline
            619784 & Cd$_{1}$Cu$_{2}$Sn$_{1}$Se$_{4}$ & 121 & strong TI & 1.7 \\
            629097 & Zn$_{1}$Cu$_{2}$Sn$_{1}$Se$_{4}$ & 121 & strong TI & 7.4 \\
            93409 & Zn$_{1}$Cu$_{2}$Ge$_{1}$Se$_{4}$ & 121 & strong TI & 10.6 \\
            629293 & Zn$_{1}$Cu$_{2}$Sn$_{1}$Te$_{4}$ & 121 & strong TI & 2.8 \\
            656153 & Zn$_{1}$Cu$_{2}$Ge$_{1}$Te$_{4}$ & 121 & strong TI & no full gap \\
            656158 & Hg$_{1}$Cu$_{2}$Sn$_{1}$Te$_{4}$ & 121 & strong TI & no full gap \\
			250247 & In$_{2}$Cu$_{2}$Se$_{4}$ & 1 & strong TI & 5.8 \\  
			674921 & Ru$_{4}$H$_{24}$ & 6 & strong TI & 23.6 \\ 
            160882 & Cu$_{2}$SnTe$_{3}$ & 44 & strong TI & no full gap \\ 
            161837 & Cd$_{2}$O$_{2}$ & 186 & strong TI & 5.4 \\ 
             
			\hline
			\hline
		\end{tabular*}
	\end{center}
 \label{table4}
\end{table*}

\section*{Discussion} 
A key material feature absent from our model is the atomic forces and stresses, which have been shown to significantly enhance the prediction of formation energy~\cite{janosh2023}. This omission likely accounts for XBERT's suboptimal performance in this task. Future research could focus on incorporating this information into the model and evaluating its impact on other complex physical properties. Another crucial yet missing aspect is the magnetic properties of materials. While magnetic properties intuitively have a considerable influence on many physical properties, the challenge lies in the complexity of magnetic structures and the lack of relevant data~\cite{Elcoro2021,Choudhary2020}. Additionally, if we identify the same material with different magnetic orderings as different samples, we should reconsider pretraining tasks to better direct the model toward learning magnetic properties.

Predicting vector properties, such as the density of states and the band structure, rather than just a single scalar property, presents a more interesting and challenging task and has been the focus of recent studies~\cite{bai2023,kong2022,li2022,li2024,chen2021}. In Supplementary Materials, we discuss combining XBERT with the recently developed Xtal2DOS model~\cite{bai2023} to make predictions on density of states. We believe that band structure can be predicted in a similar way, albeit with greater difficulty, since the band structure is much more elaborate. Such a model would be highly valuable for the material science and condensed matter physics community, given the computational intensity of calculating band structures~\cite{giustino2017}. We leave this for future work.

\section*{Methods}
\paragraph*{Datasets for pretraining and downstream tasks} For the pretraining of XBERT, we acquired data of crystal materials from the Materials Project database. Our pretraining dataset comprises 153,224 distinct materials, each characteried by its chemical formula, space group, lattice constants and Crystallographic Information File (CIF) structure. We counted the distribution of crystal systems and elements of these 153,224 materials, as shown in Supplementary Fig.S1.

\begin{table*}[htbp]
	\caption{\textbf{Overview of the datasets used for downstream tasks.} We predict the properties on 11 different datasets, 8 of which are from the Matbench suite. SC stands for superconducting.}
    \begin{center}
		\renewcommand{\arraystretch}{0.8}
		\begin{tabular*}{5.5in}
			{@{\extracolsep{\fill}}c|ccc}
			\hline
			\hline
			Dataset & Number of Samples & Property & Unit\\
			\hline
			TopoA & 9,098 & Topological Type & —— \\
            TopoB & 38,184 & Topological Type & —— \\
            3DSC  & 2,336 & SC Transition Temperature &  K \\
            Perovskites & 18,928 & Formation Energy & eV per atom \\
            Dielectric & 4,763 & Refractive Index & —— \\
            GVRH & 10,987 & Shear Modulus & log$_{10}$GPa \\
            JDFT2D & 636 & Exfoliation Energy & meV per atom \\
            KVRH & 10,987 & Bulk Modulus & log$_{10}$GPa \\
            Band Gap & 106,113 & Band Gap & eV \\
            Phonons & 1,265 & Last Phdos Peak & 1 per cm \\
            E-Form & 132,752 &Formation Energy & eV per atom\\
			\hline
			\hline
		\end{tabular*}
	\end{center}
 \label{table5}
\end{table*}

For downstream tasks, we constructed 11 datasets to test the performance of our XBERT. As shown in Table.5, we provide the dataset names, the number of samples, the properties to be predicted, and the units of the physical quantities of all these tasks.
For dataset TopoA, we adopt the non-magnetic 3D topological material database developed in Ref.~\cite{Tang2019_2}, which is available on \href{https://ccmp.nju.edu.cn}{ccmp.nju.edu.cn}. The method described in Ref.~\cite{ma2023} and Ref.~\cite{xu2024} are utilized to process the data. In brief, we combine `Topological insulators', `Topological crystalline insulators' and `Topological (semi-)metals' together into a signle category denoted as NAI (Not an Atomic Insulator). Concurrently, we denote `Materials with band crossings in case 1' as USI (Undiagnosable by Symmetry Indicators). We further divide NAI into NAI-T-IG and NAI-NT-IG, and divide USI into USI-T-IG and USI-NT-IG, based on their respective symmetry indicator groups. NAI-NT-IG and USI-NT-IG together construct the dataset we used, wherein NAI-NT-IG is classified as topological and USI-NT-IG as trivial. See Ref.~\cite{ma2023,xu2024} for a more detailed description of the procedure.
TopoB uses materials identified in Ref.~\cite{Bradlyn2017,Vergniory2019,Maia2022}, and these data are available on website \href{https://topologicalquantumchemistry.org}{topologicalquantumchemistry.org}. Materials marked as LCEBR (NLC, SEBR, ES and ESFD) are trivial (topological). We use all 38,184 materials in this database without any deletions.
The dataset for predicting superconducting transition temperature are from Ref.~\cite{Sommer2023}. In this dataset, there are 5,774 materials with recorded superconducting transition temperatures. We retained only the undoped materials, resulting in a subset of 2,336 materials.
The remaining eight datasets are all from Matbench suite~\cite{dunn2020}. In each dataset, we use all available data without any reduction or filtering.

\paragraph*{Training details} We used PyTorch to generate data, build the model, and execute the training process. To obtain the conclusions in Table.1, we employed a network architecture comprising 3 layers of graph convolution and 5 layers of encoder (with 4 attention heads). For the purpose of model interpretability, we computed $\Bar{s}_{sg}$ and $\sum_{j}|\omega_{ij}|$ utilizing a simplified architecture of 1 layer and 1 attention head in the encoder layer. This reduction in complexity was necessary to mitigate the conflation of information from disparate tokens that occurs with increased layer depth. Across both pre-training and downstream tasks, we set the batch size to be 64. For regression tasks, we used the mean squared error (MSE) as the loss function, while for classification tasks, we used the cross-entropy loss function. We consistently employed the AdamW optimizer, with the learning rate, weight decay and epoch size detailed in Supplementary Materials. 

We employed $k$-fold cross-validation cross-validation to evaluate XBERT's performance across various tasks. The complete dataset is initially subjected to random shuffling and subsequently partitioned into $k$ mutually exclusive subsets of approximately equal size. In each iteration of the $k$-fold process, one subset is designated as the validation set, while the amalgamation of the remaining $k-1$ subsets constitutes the training set. The model, after being trained on the designated training set, is then evaluated on the held-out validation set to compute performance metrics. This process is iterated $k$ times, with each subset serving once as the validation set.  Finally, we average the results of the $k$ training-validation cycles to obtain the final result. Specifically, TopoA used 11-fold cross-validation (to be consistent with Ref.~\cite{ma2023}), TopoB used 10-fold cross-validation (to be consistent with Ref.~\cite{claussen2020}), and the remaining tasks used 5-fold cross-validation.

In the evaluation of binary classification tasks, many evaluation metrics are employed, including accuracy, recall, precision, and the F1 score. To elucidate these metrics we first use TP to denote the number of true positives (i.e., number of samples that are actually nontrivial and predicted to be nontrivial), FP to denote the number of false positives (i.e., number of samples that are actually trivial but predicted to be nontrivial), TN to denote the number of true negatives (i.e., number of samples that are actually trivial and predicted to be trivial), and FN to denote the number of false negatives (i.e., number of samples that are actually nontrivial but predicted to be trivial). Then we have
\begin{equation}
	\mathrm{accuracy = \frac{TP+TN}{TP+FP+FN+TN}}
\end{equation}
\begin{equation}
	\mathrm{precision = \frac{TP}{TP+FP}}
\end{equation}
\begin{equation}
    \mathrm{recall = \frac{TP}{TP+FN}}
\end{equation}
\begin{equation}
	\mathrm{F_{1}\ score = \frac{2 \cdot precision \cdot recall}{precision + recall}}
\end{equation}

\paragraph*{Details on First Principles Calculations} The first principles calculations are carried out in the framework of the generalized gradient approximation (GGA) functional of the density functional theory through employing the Vienna ab initio simulation package (VASP) with
projector augmented wave pseudopotentials~\cite{Kresse1996,Bl1994,Perdew1996}. The convergence criterion for the total energy is $10^{-6}$ eV and the lattice constants and inner positions are obtained through full relaxation with a force tolerance criterion for convergence of 0.01~eV/\AA. The $k$-point meshes of $8\times 8 \times 8$ are used to sample the Brillouin zones. The SOC effect is self-consistently included. The Hubbard $U$ correction is introducted by LDA+U functional~\cite{Dudarev1998}. Surface state calculations, namely the LDOS and Wannier charge center are performed based on maximally localized Wannier functions (MLWF) by Wannier90~\cite{Arash2008} and the WannierTools packages~\cite{QuanSheng2018}. 

\newpage


\newpage

\begin{addendum}
\item We thank Xiao-Liang Qi for enlightening discussions. This work is supported by the National Key Research Program of China under Grant No.~2019YFA0308404, the Natural Science Foundation of China through Grants No.~12350404 and No.~12174066, the Innovation Program for Quantum Science and Technology through Grant No.~2021ZD0302600, the Science and Technology Commission of Shanghai Municipality under Grants No.~23JC1400600 and No.~2019SHZDZX01. All data needed to evaluate the conclusions in the paper are present in the paper and/or the Supplementary Materials.
\item[Author contribution] J.W. supervised the project. D.Q. designed XBERT's structure. H.X. trained the XBERT model and performed the DFT calculations. All authors contributed to designing the research, interpreted the results, and wrote the manuscript.
\item[Competing interests] The authors declare that they have no competing interests.
\end{addendum}

\newpage
\begin{figure}[t]
\begin{center}
\includegraphics[width=5.4in, clip=true]{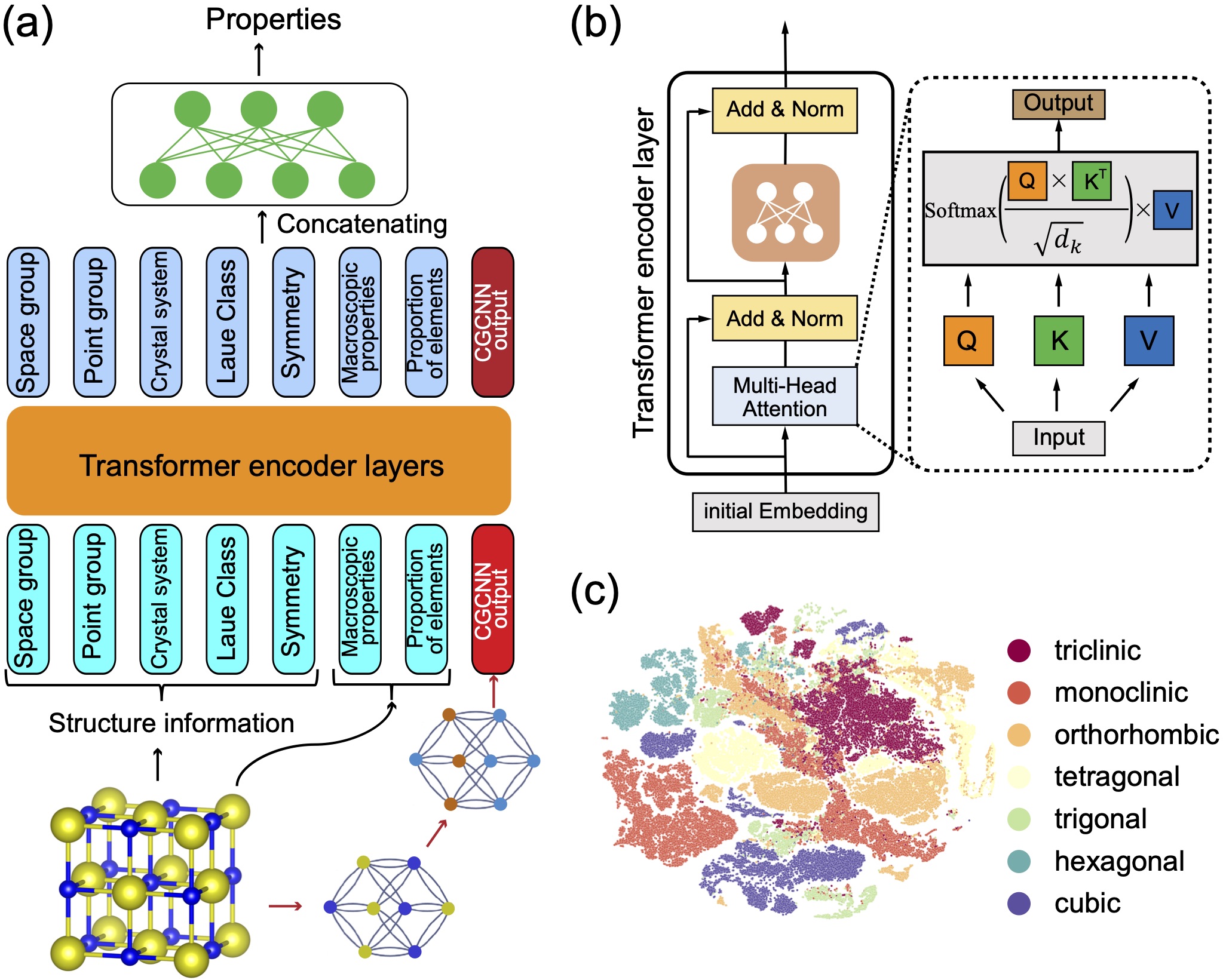}
\end{center}
\caption{XBERT's structure and t-SNE result. (a) XBERT's structure. A crystalline material's structural and elemental information are encoded into eight tokens. The last token is derived from the output of a CGCNN, which captures unit cell information. These tokens are then fed into the transformer encoder and a single fully connected layer is applied to the output tokens with outputs depending on the specific task. (b) The detailed structure of a transformer encoder layer. We choose “Scaled Dot-Product Attention” with queries, keys and values are from different embeddings of the input vector. 
(c) 2D representation of the crystal feature vector by t-SNE.}
\label{fig1}
\end{figure}

\begin{figure}[b]
\begin{center}
\includegraphics[width=5.4in, clip=true]{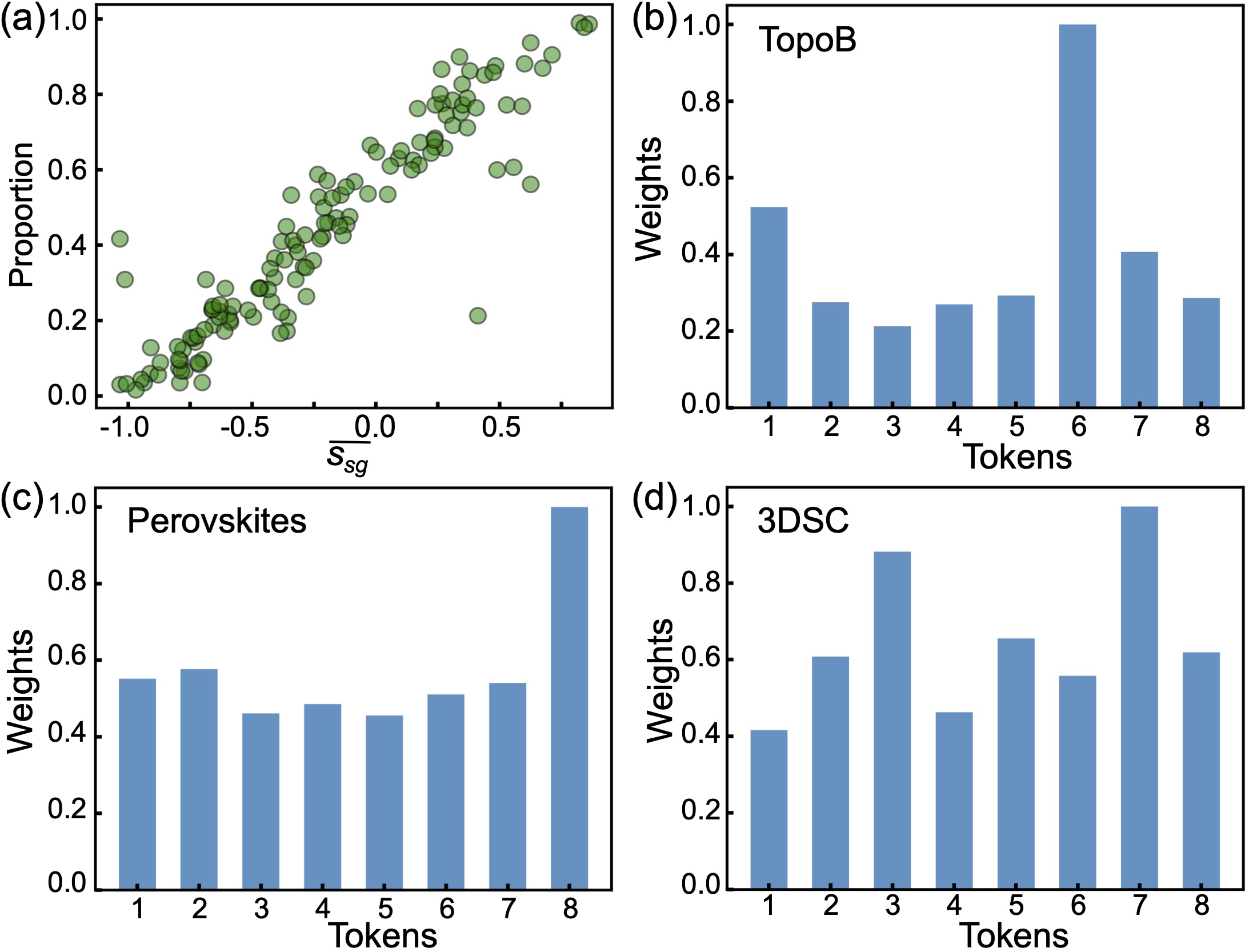}
\end{center}
\caption{Interpretation of the model. (a) The relationship between the mean space group feature value $\overline{s_{sg}}$ and the proportion of topological materials within each space group. Each point represents a specific space group, exhibiting a clear linear correlation. This suggests that the output token reflects the frequency of topological materials within that space group to a significant extent.(b)-(d) Bar charts of weights $\sum_j{\left | w_{ij} \right |}$ assigned to different features for predicting topological property, formation energy and superconducting transition temperature. It is important to note that tokens 1-5 capture space group information, tokens 6-7 capture elemental information, and token 8 captures unit cell information.}
\label{fig2}
\end{figure}

\begin{figure*}[htbp]
\begin{center}
\includegraphics[width=6.4in, clip=true]{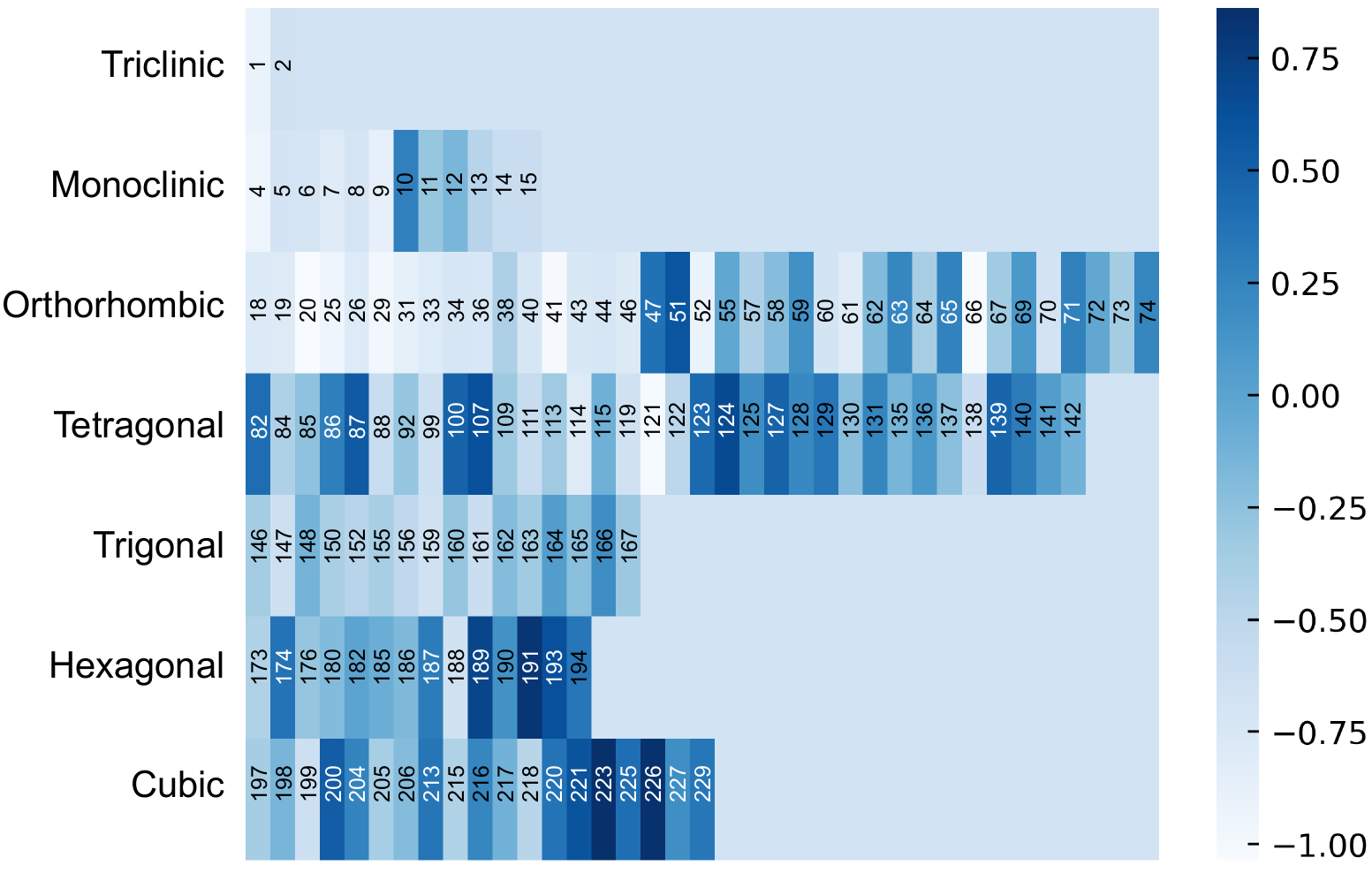}
\end{center}
\caption{The visualization diagram of $\Bar{s}_{sg}$. The numbers displayed represent space group numbers. It should be noted that certain space groups are not present due to the absence of corresponding materials in the TopoA database. Within a given crystal system, space groups with higher group numbers generally exhibit a greater number of symmetry operations and consequently possess higher overall symmetry.}
\label{fig3}
\end{figure*}

\begin{figure}[t]
\begin{center}
\includegraphics[width=5.4in, clip=true]{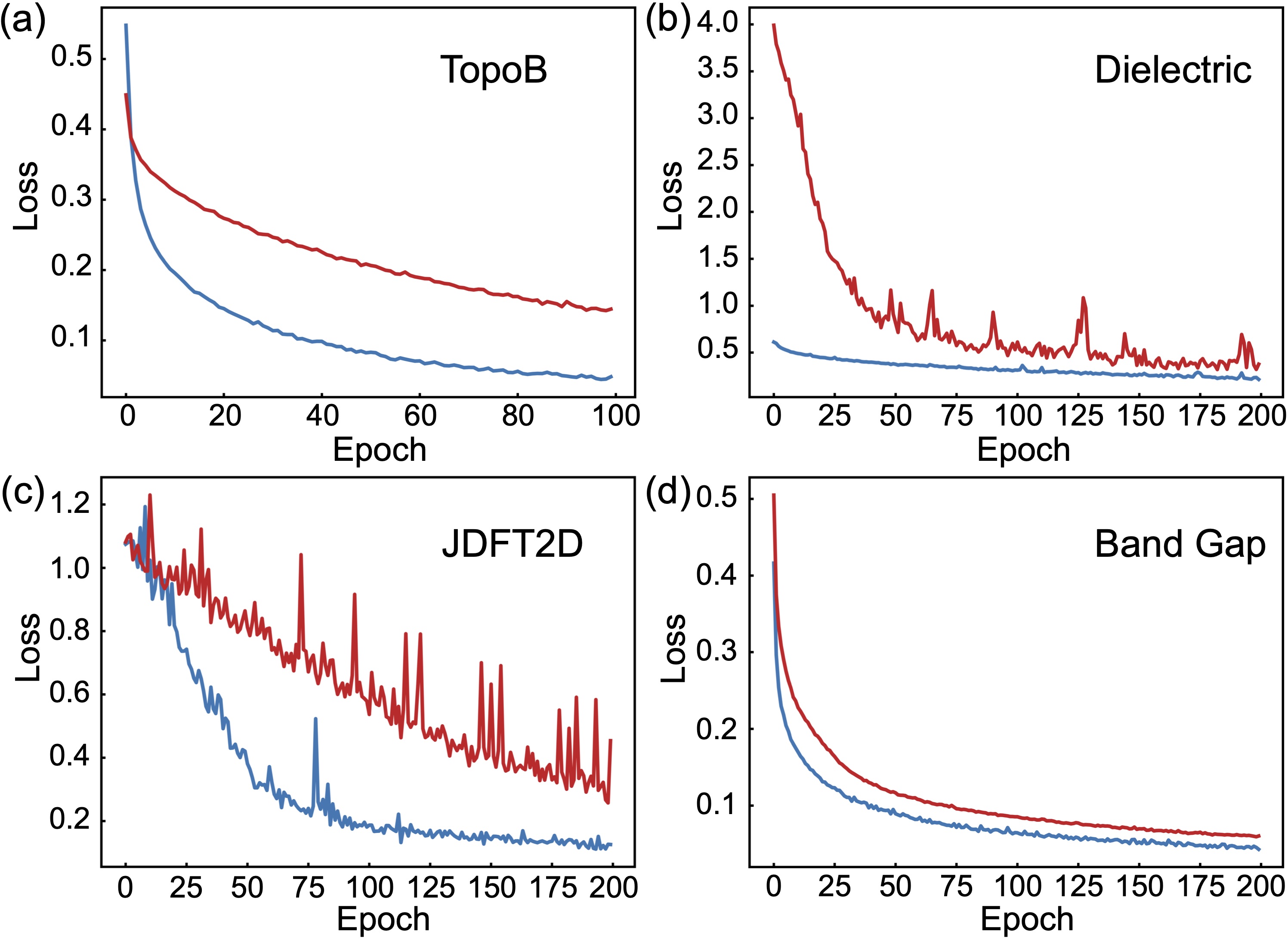}
\end{center}
\caption{Efficiency of XBERT. We selected four tasks: (a) topological property, (b) dielectric constant, (c) exfoliation energy, and (d) electronic band gap, and compared the reduction in training loss over time. The red and blue lines correspond to CGCNN and XBERT, respectively. XBERT's training loss converges in fewer epochs compared to CGCNN.}
\label{fig4}
\end{figure}

\begin{figure}[b]
\begin{center}
\includegraphics[width=5.4in, clip=true]{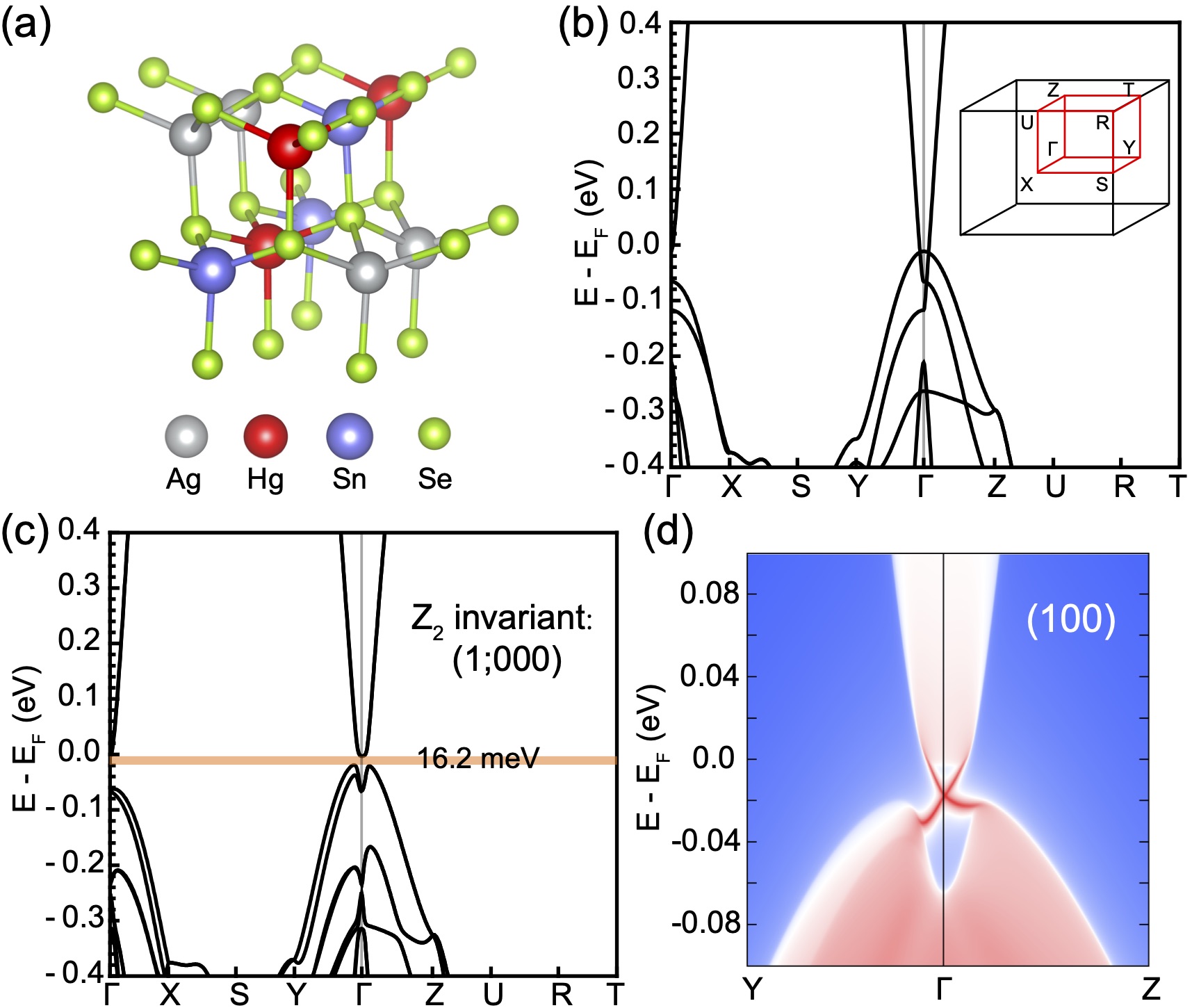}
\end{center}
\caption{Electronic structure and topological property of Ag$_{2}$HgSe$_{4}$Sn. (a) Crystal structure with space group $Pmn2_1$ (No.~31). (b), (c) Band structure without and with SOC, the $Z_2$ invariant is $(1;000)$~\cite{fu2007}. (d) A single Dirac cone surface state.}
\label{fig5}
\end{figure}

\end{document}



\baselineskip24pt


\maketitle

\textbf{This PDF file includes:}

Supplementary Text

Tables. S1 to S4

Figs. S1 to S12


\newpage

\section{The distritution of pertraining dataset.}

The distribution of crystal systems and elements of the 153,224 materials we used for pertraining are shown in Fig.S1.

\section{Details on transformer encoder layer.}

To comprehend the Transformer encoder layer~({\it 35, 46\/}), let's first consider a simple example: a material described by 8 tokens(each represented by a 64-dimension vector) passing through a single-head, single-layer encoder. As shown in Fig.~1(b), the encoder layer comprises two parts: 1)~scaled dot-product self-attention calculation, residual connection and layer normalization; 2)~fully connected layer, residual connection and layer normalization. We will mainly focus on the self-attention mechanism, as the rest is straightforward. Initially, this material can be represented by an $8\times 64$ matrix. By multiplying this matrix with three different $64\times 64$ matrices, we obtain three different $8\times 64$ matrices, denoted as $Q$, $K$, and $V$, representing the queries, keys, and values matrices, respectively. Next, by multiplying $Q$ with the transpose of $K$ and applying the Softmax activation function, we obtain the attention scores matrix. Multiplying this with the $V$ matrix yields the updated representing matrix $Z$. The entire process can be expressed mathematically as:
    $\text{Attention}\left(Q,K,V\right) = \text{softmax}\left(\frac{QK^{T}}{\sqrt{d_{k}}}\right)V$.
To gain insight into this computation, let's denote the rows of $(Q,K,V)$ as $(q_{1},k_{1},v_{1})$, $(q_{2},k_{2},v_{2})$,...,$(q_{8},k_{8},v_{8})$. The first row of the attention score matrix is $[S(q_{1}\cdot k_{1}),S(q_{1}\cdot k_{2}),...,S(q_{1}\cdot k_{8})]$, which can be interpreted as each token's contribution to the first one. (Here, $\cdot$ means scalar product and $S$ denotes the Softmax function. We omit $d_{k}$ here for simplicity, as it is merely a normalizing constant.) Furthermore, the first row of $Z$ (i.e., the updated vector describing the first token) can be succinctly written as $S(q_{1}\cdot k_{1})v_{1}+S(q_{1}\cdot k_{2})v_{2}+......+S(q_{1}\cdot k_{8})v_{8}$. This implies that if the $i$-th token contributes significantly to the first token (i.e., $S(q_{1}\cdot k_{i})$ is large), then $v_{i}$ will have a greater influence during the updating procedure. Lastly, we briefly discuss the concept of mask in Transformers. To disregard the contribution of the $j$-th token to the first one, we can simply set $q_{1}\cdot k_{j}$ to negative infinity. Consequently, after applying Softmax, this term becomes zero, and $v_{i}$ will no longer contribute to the first token. In this work, we did not utilize masks. However, if we want to handle tokens of unequal lengths, masks will be necessary.

We now turn to multi-head attention mechanisms. Consider a batch of materials passing through a 4-head self-attention calculation layer. Let us assume the batch size is 64. Initially, each batch can be represented by a $64\times 8\times 256$ tensor, which can be conceptualized as 64 matrices, each with 8 rows and 256 columns. The dimensions correspond to the following: 64 represents the number of materials in a batch, 8 denotes the number of tokens describing each material, and 256 signifies the dimensionality of each token's vector representation. For the multi-head attention calculation with 4 heads, we first reshape the tensor into the shape of $64\times 4 \times 8 \times 64$. This can be interpreted as 64 stacks of matrices, with each stack comprising 4 matrices of size $8\times 64$. In this new configuration, 64 still corresponds to the batch size, 4 represents the number of attention heads, 8 denotes the number of tokens per material, and 64 signifies the reduced dimensionality of each token's vector representation. Subsequently, for each $8\times 64$ matrices, we can perform the single-head attention calculation as introduced before. Upon completion of these calculations, we reshape the resultant tensor back to its original dimensions of $64\times 8 \times 256$. This process constitutes the complete multi-head attention calculation.

\section{hyperparameters}
For the XBERT$_\text{FC}$-based models, we used a fixed learning rate, with the learning rate, weight decay and epoch size for different tasks listed in Table ~\ref{hyperparamater}. For the XBERT$_\text{FA}$ model, we adopted the one-cycle policy with the maximum learning rate, weight decay and epoch size listed in Table ~\ref{hyperparamater2}.

\begin{table*}[htbp]
	\caption{\textbf{Hyperparamter for different tasks in XBERT$_\text{FC}$-based models}}
    \begin{center}\label{results2}
		\renewcommand{\arraystretch}{2}
		\begin{tabular*}{4in}
			{@{\extracolsep{\fill}}c|ccc}
			\hline
			\hline
			Tasks & Learning Rate & Weight Decay& Epoch Size \\
			\hline
            Pertrain & 0.0001 & 1e-8 & 200 \\
            \hline
			TopoA & 0.0001 & 1e-8 & 100\\
            TopoB & 0.0001 & 1e-8 & 100\\
            3DSC  & 0.0001 & 0.0001 & 100\\
            Perovskites & 0.00001 & 0.01 & 200\\
            Dielectric & 0.0001 & 0.001 & 200\\
            GVRH & 0.0001 & 0.01 & 200 \\
            JDFT2D & 0.00001 & 0.001 & 200\\
            KVRH & 0.0001 & 0.001 & 200\\
            Band Gap & 0.0001 & 0.0001 & 200\\
            Phonons & 0.0001 & 0.001 & 200\\
            E-Form & 0.0001 & 0.0001 & 200\\
			\hline
			\hline
		\end{tabular*}
	\end{center}
 \label{hyperparamater}
\end{table*}

\begin{table*}[htbp]
	\caption{\textbf{Hyperparamter for different tasks in XBERT$_\text{FA}$ model}}
    \begin{center}\label{results2}
		\renewcommand{\arraystretch}{2}
		\begin{tabular*}{4in}
			{@{\extracolsep{\fill}}c|ccc}
			\hline
			\hline
			Tasks & Learning Rate & Weight Decay& Epoch Size \\
           \hline
			TopoA & 0.001 & 1e-8 & 100\\
            TopoB & 0.001 & 1e-8 & 100\\
            3DSC  & 0.0001 & 0.0001 & 300\\
            Perovskites & 0.001 & 0.01 & 300\\
            Dielectric & 0.0001 & 0.001 & 300\\
            GVRH & 0.001 & 0.01 & 300 \\
            JDFT2D & 0.0001 & 0.001 & 300\\
            KVRH & 0.001 & 0.001 & 300\\
            Band Gap & 0.001 & 0.0001 & 300\\
            Phonons & 0.001 & 0.001 & 300\\
            E-Form & 0.001 & 0.0001 & 300\\
			\hline
			\hline
		\end{tabular*}
	\end{center}
 \label{hyperparamater2}
\end{table*}

\section{Further analysis of the performance}
\subsection{Effect of normalization}
In the main text, models V and VI represent the unnormalized versions of models XBERT$_{F}$ and II, respectively. The normalization process involves applying a Softmax function to each of the 256-dimensional vectors obtained from the encoder layer, ensuring that the sum of the absolute values of each vector's 256 components equals 1. This process aims to enhance the interpretability of $\sum_{j}|w_{ij}|$. However, it is noteworthy that for certain tasks where CGCNN demonstrates high efficacy, the unnormalized versions tend to outperform the normalized ones. A notable example is the Phonons task. To understand this phenomenon, we randomly selected 8 materials and output their 256-dimensional vectors post-encoder layer using model V, after fine-tuning with the Phonons task. We then calculated the sum of the absolute values of the 256 components for each vector. The results are presented in Table~\ref{sum_of_abs}, where the maximum and second-largest values in each row are highlighted in red and blue, respectively. It is evident that for tasks where CGCNN excels, the sum of the absolute values of the vectors output by CGCNN tends to be larger. The normalization process, by design, would negate this effect, potentially leading to a reduction in predictive accuracy.

\begin{table*}[htbp]
	\caption{\textbf{The sum of the absolute values of the 256 components for each token.} The largest (second largest) value is bolded (underlined). SC, PG, CS, LC, NE, and FE stand for space group, point group, crystal system, Laue class, number of electrons, and elemental fraction, respectively.}
    \begin{center}\label{results3}
		\renewcommand{\arraystretch}{2}
		\begin{tabular*}{6in}
			{@{\extracolsep{\fill}}cccccccc}
			\hline
			\hline
			SG & PG & CS & LC & Symmetry & NE & FE & CGCNN\\
			\hline
            161.56 & 166.03 & 164.31 & 158.40 & 169.20 & 160.32 & \underline{169.39} & \textbf{171.70}\\
            \textbf{174.01} & 164.96 & 164.96 & 164.96 & 169.67 & 162.50 & 163.51 & \underline{172.88}\\
            163.40 & 158.12 & 158.94 & 163.99 & \underline{165.55} & 163.66 & 159.41 & \textbf{172.70}\\
            165.13 & 160.55 & 152.84 & 160.55 & 160.94 & 153.03 & \textbf{174.76} & \underline{169.45}\\
            \textbf{161.47} & 153.03 & \underline{158.47} & 153.03 & 158.17 & 157.38 & 156.94 & 158.26\\
            168.66 & 169.86 & 154.60 & 169.86 & 165.09 & 159.66 & \textbf{173.64} & \underline{173.48}\\
            172.18 & 166.69 & 160.89 & 166.69 & \underline{173.16} & 161.64 & 160.63 & \textbf{175.60}\\
            163.15 & 161.39 & 156.75 & 161.39 & \underline{166.10} & 157.33 & 165.14 & \textbf{168.44}\\
			\hline
			\hline
		\end{tabular*}
	\end{center}
 \label{sum_of_abs}
\end{table*}

\subsection{Attention weights}

Fig.S2-S5 present the attention weights. Fig.S2 and S3 (S4 and S5) show the results for the XBERT$_{F}$(II) model. Fig.S2 and S4 display the results after pre-training, whereas Fig.S3 and S5 display the results after fine-tuning on the TopoA task. Each figure labeled `Contribution to Token $i$' illustrates the contribution of different tokens to the $i$-th token in each encoder layer, with contributions from different heads summed together. These figures are generated using [AlI$_{4}$Na, SG=62] as the input material. The figures indicate that during both training and fine-tuning, each token's information is accessible to other tokens. Additionally, the attention weights after fine-tuning are not significantly different from the pre-trained attention weights.

\section{DOS prediction}
Ref.69 proposes a fully attention-based transformer decoding for density of states (DOS) prediction. We combine this model with XBERT to predict electronic DOS. 

Fig.S6 illustrates our architecture. We first use XBERT to extract the global feature of the material, which are then duplicated $n$ times (corresponding to the $n$ data points in the density of states). Each copy is augmented with positional encoding. These augmented features are then processed through $N$ encoder layers. Each encoder layer comprises three parts: the first part performs self-attention computation on the encoded global features, the second part performs source-attention computation between the global features and the representation vectors of each atom in the material, and the final part applies layer normalization, a fully connected layer, and residual connections. After the final encoder layer, each vector is mapped to a single value, resulting in $n$ data points for the density of states.

We randomly selected 50,000 materials with density of states data from the Materials Project for the training set and 8,000 materials for the test set. We focused on the region within $\pm$4 eV around the Fermi level and uniformly sampled 128 data points within this range. We used CGCNN, XBERTF, II, and VI models for prediction, and the results are presented in the Table~\ref{dos}.

We presented the results of mean absolute error (MAE) and R-square (R$^{2}$) for these three models after 50 epochs. The results indicate that XBERT's performance is slightly worse than CGCNN. This may be due to the fact that the information added by the XBERT's encoder layer is not highly relevant to the density of states (DOS), potentially leading to a slight degradation in predictive accuracy. We hypothesize that the integration of more relevant properties into the model could yield performance improvements. Furthermore, there exists potential for architectural modifications to enhance the model's efficacy. We leave this for future work.

\begin{table*}[htbp]
	\caption{\textbf{The performance of different models in the task of predicting DOS measured by MAE and R$^{2}$.}}
	\begin{center}\label{results7}
		\renewcommand{\arraystretch}{1.3}
		\begin{tabular*}{3.0in}
			{@{\extracolsep{\fill}}ccc}
			\hline
			\hline
			Model &MAE &R$^{2}$\\
			\hline
			CGCNN & 3.829 & 0.493\\  
            XBERT$_{F}$ & 3.876 & 0.478\\  
            II & 3.907 & 0.476\\  
			VI & 3.808 & 0.477\\ 
			\hline
			\hline
		\end{tabular*}
	\end{center}
 \label{dos}
\end{table*}

\section{First Principles Calculations of A\MakeLowercase{g}$_{2}$H\MakeLowercase{g}S\MakeLowercase{e}$_{4}$S\MakeLowercase{n}}

 In the main text, we performed the calculations for material Ag$_{2}$HgSe$_{4}$Sn using the structure directly downloaded from the database ccmp.nju.edu.cn (the .struct file is converted to POSCAR) without relaxing the lattice constants. Since the orbitals near the Fermi level are primarily contributed by the $p$-electrons of Se, we did not apply a Hubbard $U$ correction to the $d$-electrons of Ag and Hg atoms. However, in the supplementary materials, we also tested the effects of applying a Hubbard U correction and relaxing the lattice constants.

Fig.S7 shows the Wilson loop calculated using the standards used in the main text, clearly indicating a topological index of (1;000), which signifies a three-dimensional strong topological insulator. Fig.S8 presents the results of our calculations after applying a Hubbard U correction and relaxing the lattice constants. The results demonstrate that neither the Hubbard U correction nor the lattice constant relaxation significantly altered the band structure or the topological properties. Notably, after relaxing the lattice constants, the material's energy gap increased to 24.2 meV.

\section{More topological insulators.}
We observed that in the TopoB dataset, topological materials lacking inversion symmetry may not be identifiable by symmetry indicator and could therefore be misclassified as trivial. To address this, we divided the TopoB dataset into two parts based on the presence or absence of inversion symmetry. The dataset contains 30,733 materials with inversion symmetry and 7,451 without. Of those lacking inversion symmetry, 4,826 are originally classified as trivial.
 Then we used the 30733 materials with inversion symmetry to retrain the model and made predictions on these 4,826 materials. Approximately one-fifth of the materials were selected by our XBERT model. After excluding metals and materials with very large band gaps, we performed first-principles calculations on the remaining candidates and identified 10 topological insulators.

Band structures and Wilson loops for the ten new topological insulators are shown in Fig.S9-Fig.S12. These results were obtained by directly using the structure files available on the website. If structural relaxation is performed, the gap of some materials may decrease, but their topological properties remain unchanged.


\clearpage

\newpage

\begin{figure*}[htbp]
\begin{center}
\includegraphics[width=6.4in, clip=true]{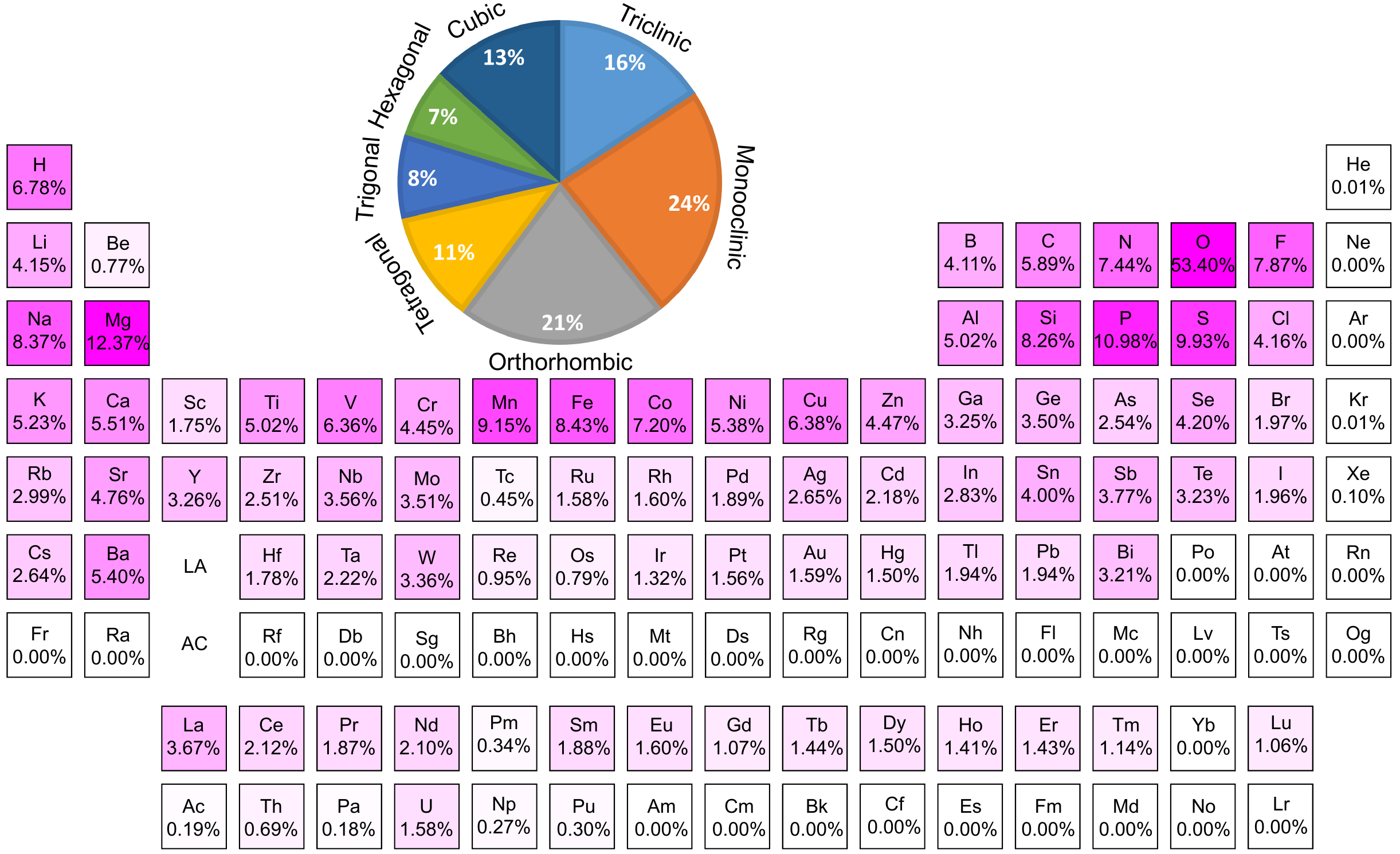}
\end{center}
\caption{The distribution of crystal systems and elements of the pretraining dataset. Pie chart: the proportion of materials belonging to different crystal systems. Periodic table: the proportion of materials containing each element.}
\label{figS0}

\end{figure*}
\begin{figure*}[htbp]
\begin{center}
\includegraphics[width=6.4in, clip=true]{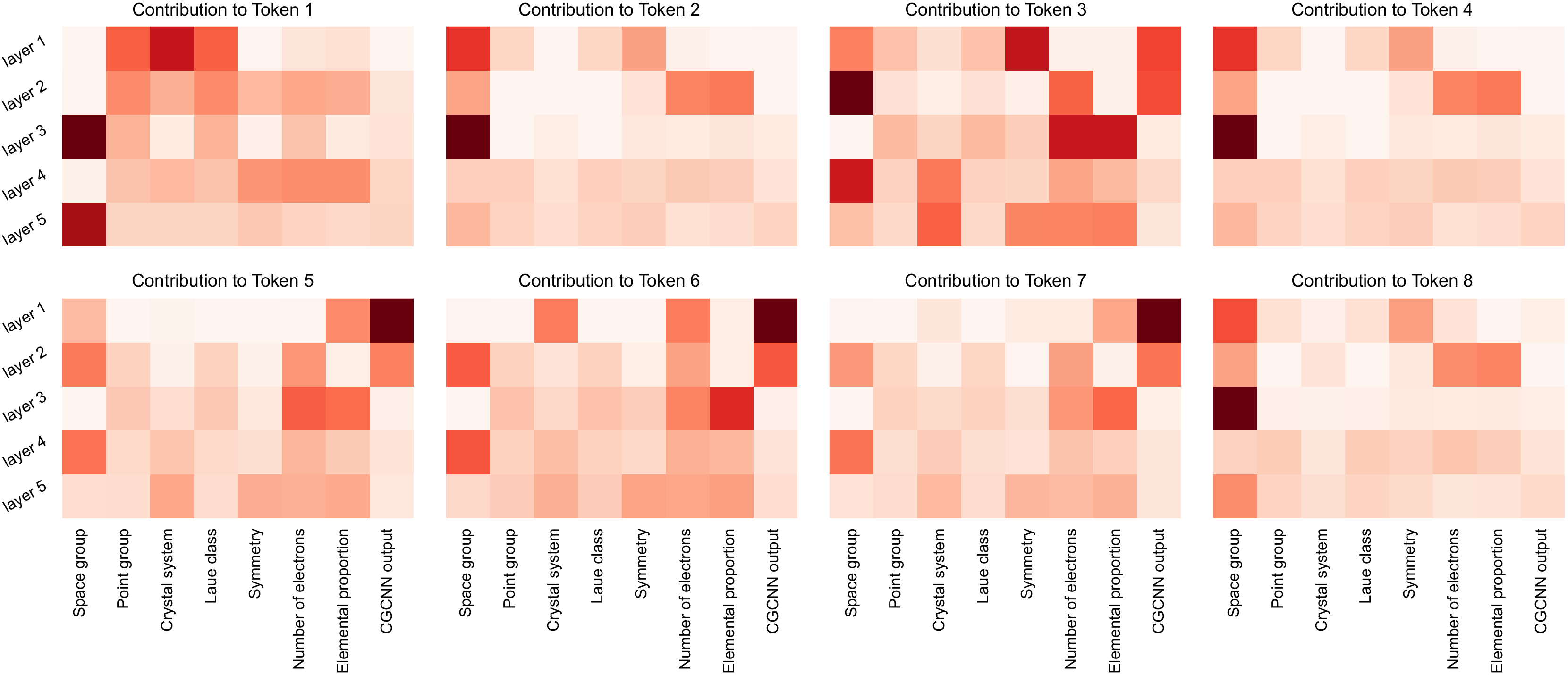}
\end{center}
\caption{The atention weights of XBERT$_{F}$ after pretraining, with the input material being [AlI$_{4}$Na, SG=62].}
\label{figS1}
\end{figure*}

\begin{figure*}[htbp]
\begin{center}
\includegraphics[width=6.4in, clip=true]{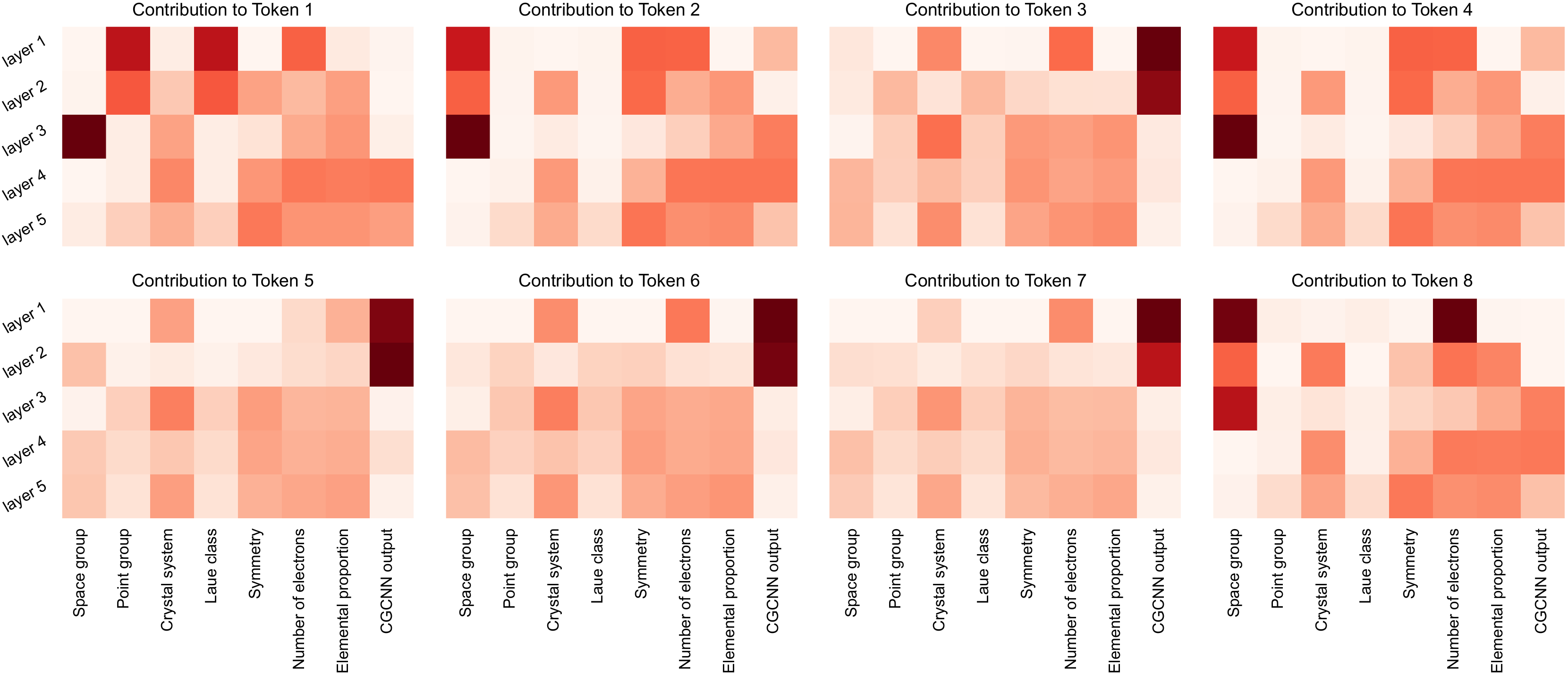}
\end{center}
\caption{The atention weights of XBERT$_{F}$, fine-tuned with the topoA task, with the input material being [AlI$_{4}$Na, SG=62].}
\label{figS2}
\end{figure*}

\begin{figure*}[htbp]
\begin{center}
\includegraphics[width=6.4in, clip=true]{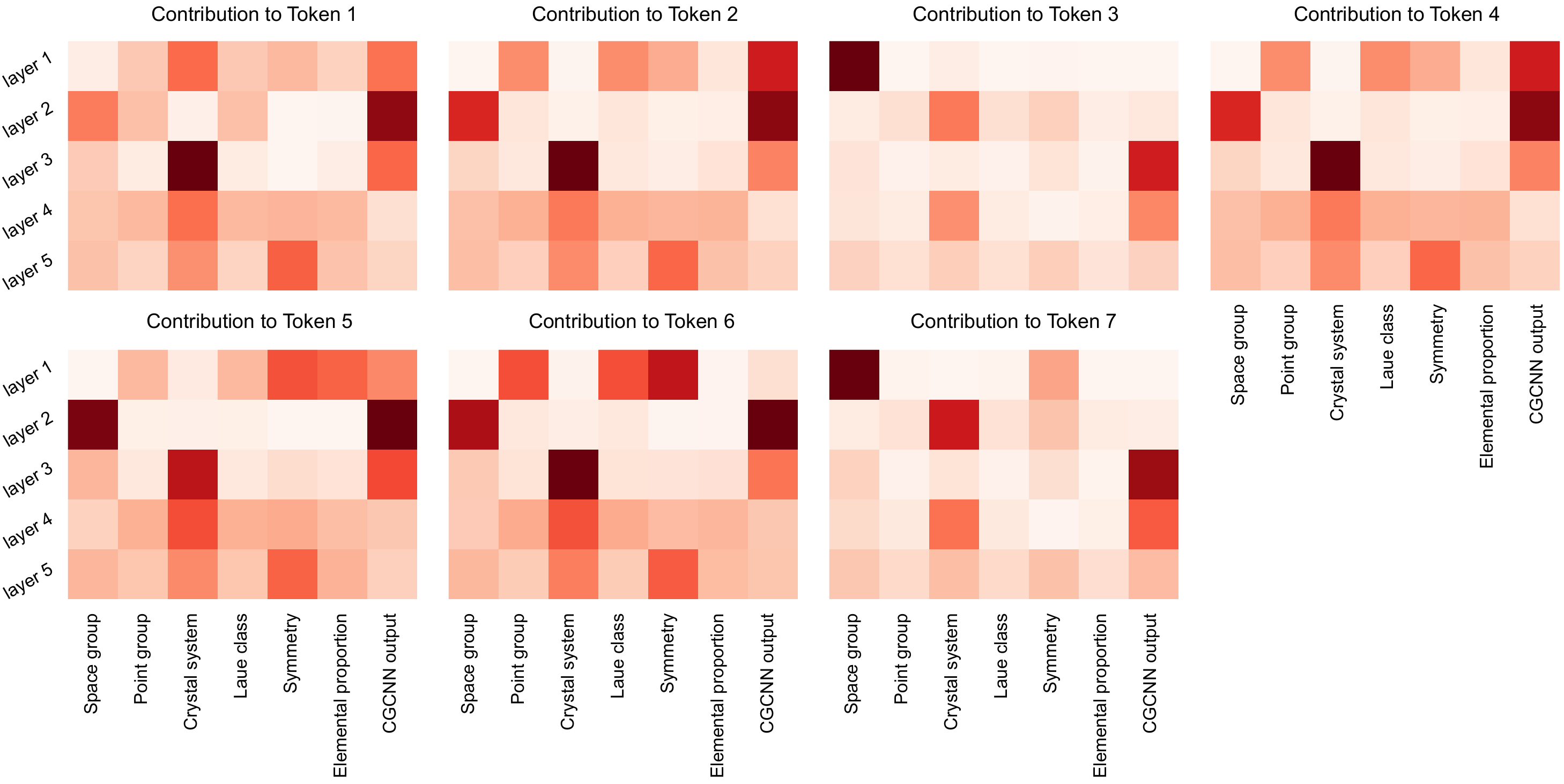}
\end{center}
\caption{The atention weights of model II after pretraining, with the input material being [AlI$_{4}$Na, SG=62].}
\label{figS3}
\end{figure*}

\begin{figure*}[htbp]
\begin{center}
\includegraphics[width=6.4in, clip=true]{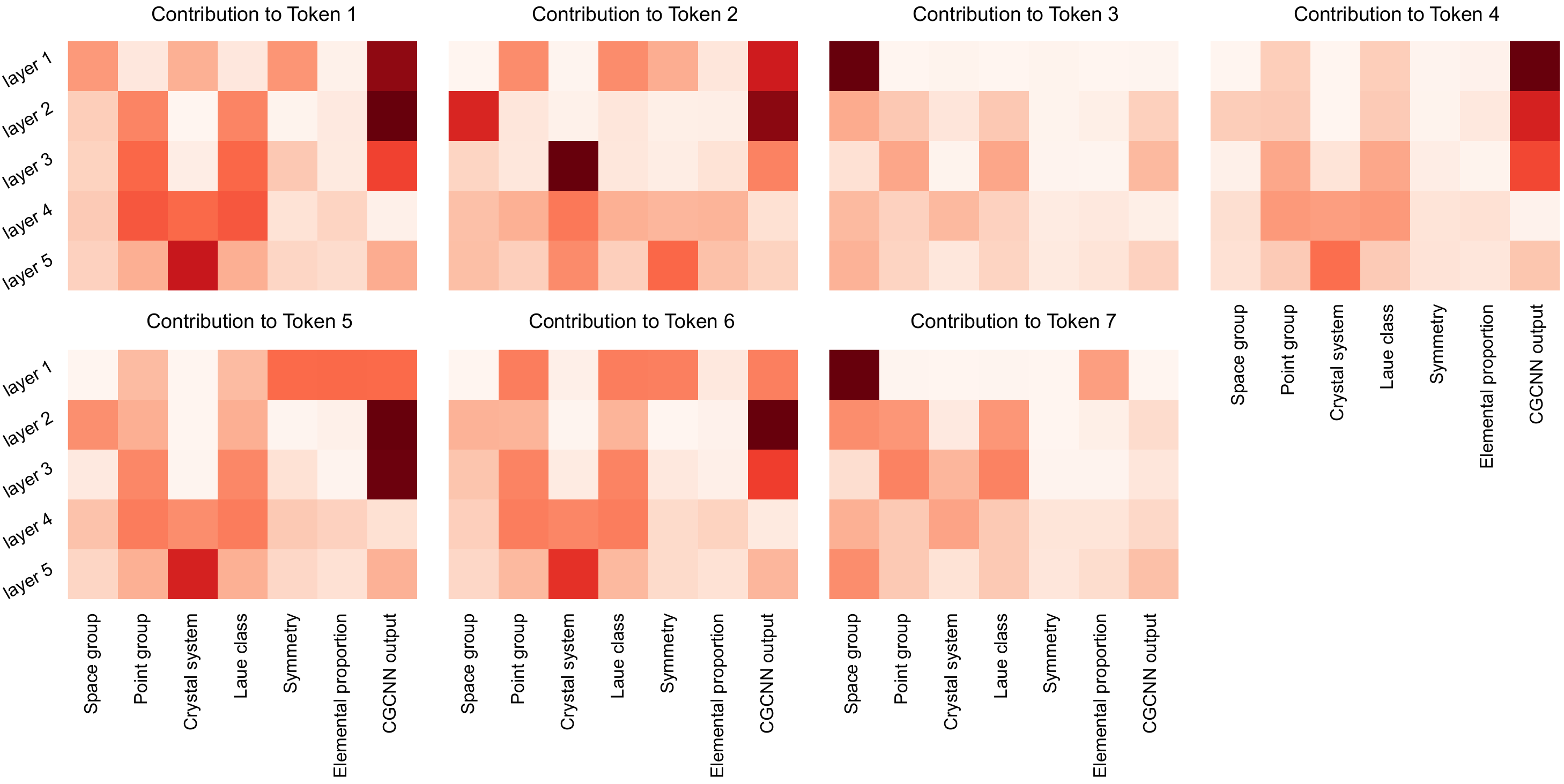}
\end{center}
\caption{The atention weights of model II, fine-tuned with the topoA task, with the input material being [AlI$_{4}$Na, SG=62].}
\label{figS4}
\end{figure*}

\begin{figure*}[htbp]
\begin{center}
\includegraphics[width=6.4in, clip=true]{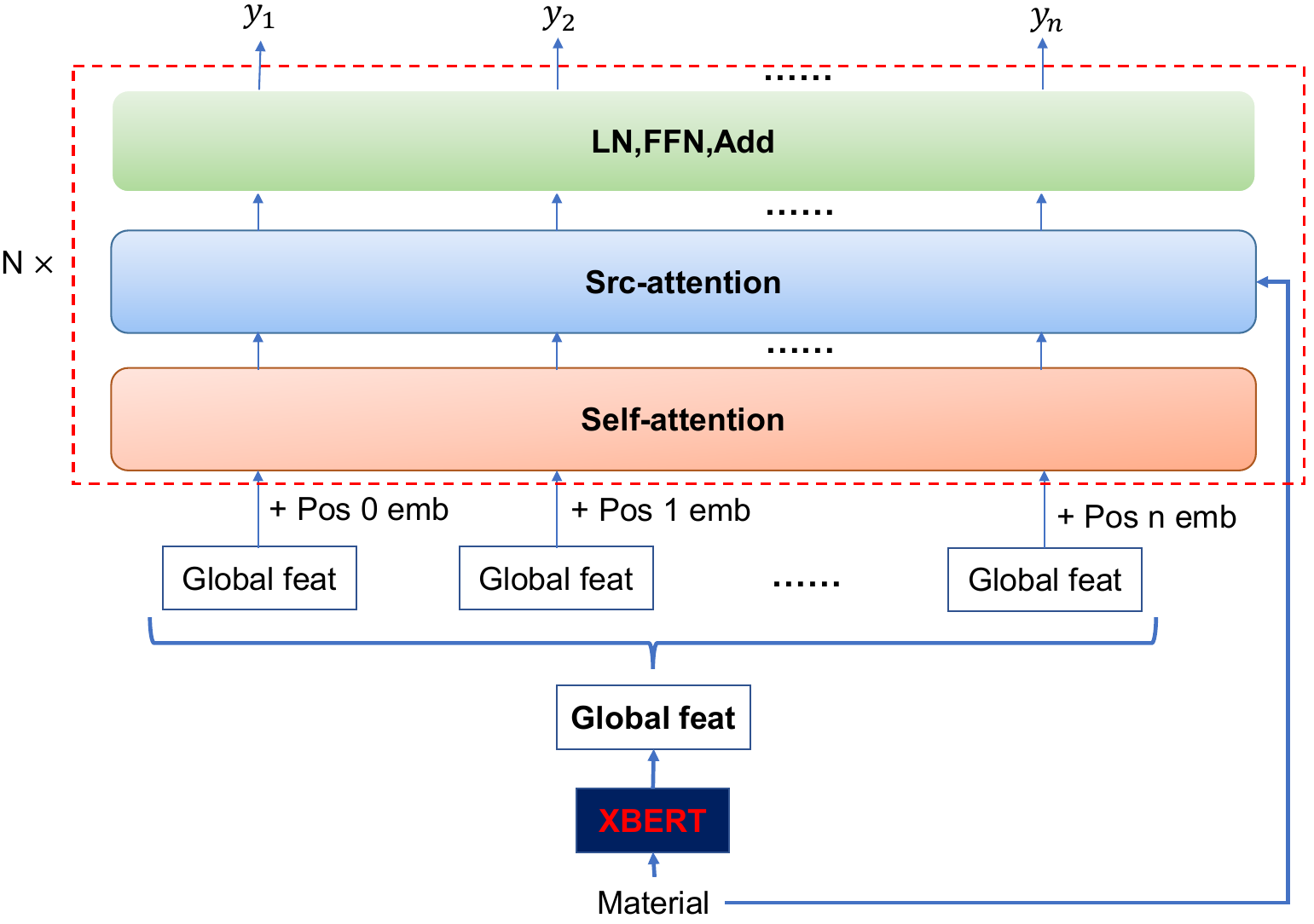}
\end{center}
\caption{The architecture for predicting DOS. LN denotes Layer Normalization, FFN denotes Feedforward Network, and Add denotes Addition.}
\label{figS5}
\end{figure*}

\begin{figure*}[htbp]
\begin{center}
\includegraphics[width=4in, clip=true]{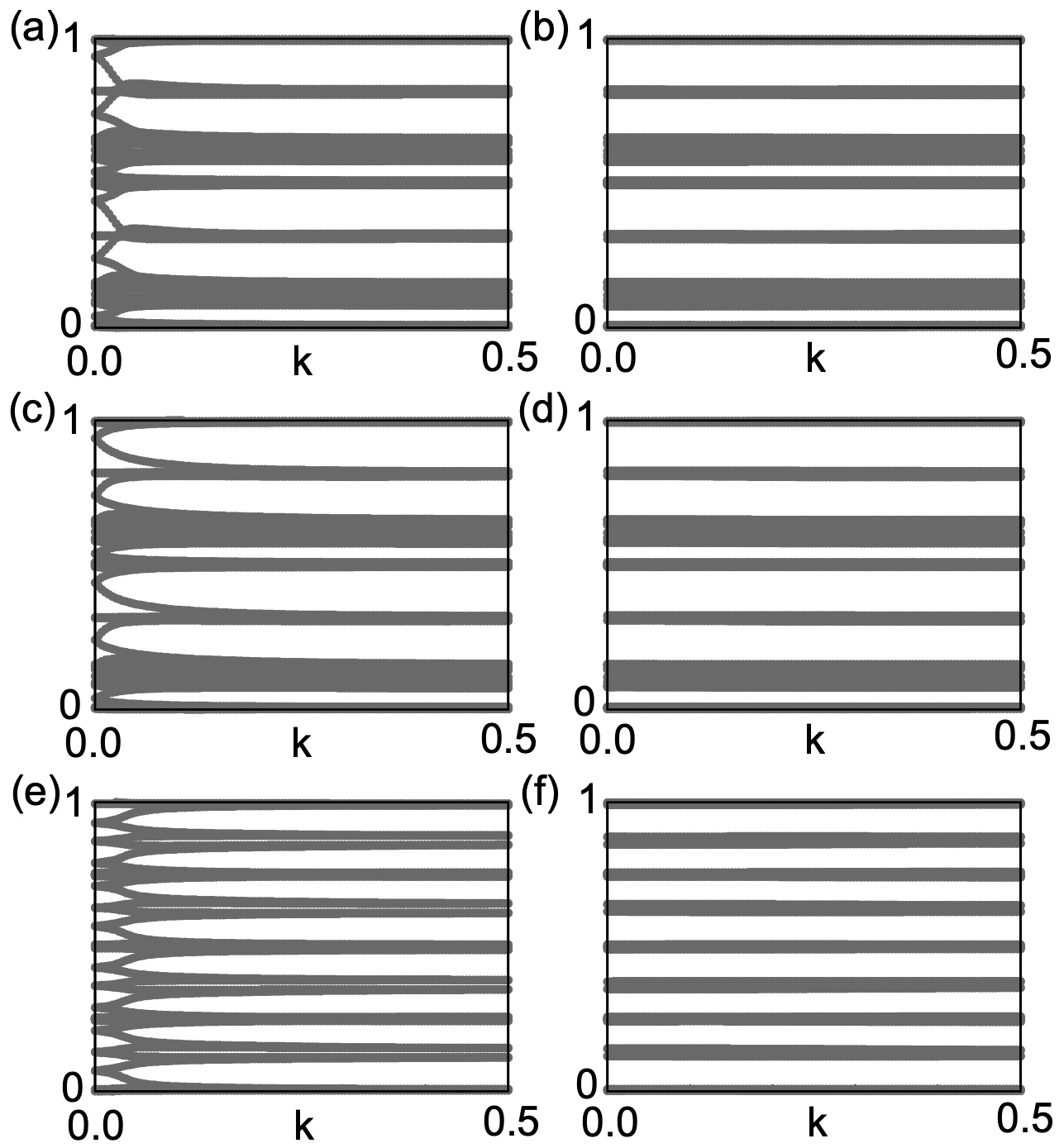}
\end{center}
\caption{The Wilson loop for Ag$_{2}$HgSe$_{4}$Sn. (a) $k_{1}=0, k_{2}-k_{3}$ plane. (b) $k_{1}=0.5, k_{2}-k_{3}$ plane. (c) $k_{2}=0, k_{1}-k_{3}$ plane. (d) $k_{2}=0.5, k_{1}-k_{3}$ plane. (e) $k_{3}=0, k_{1}-k_{2}$ plane. (f) $k_{3}=0.5, k_{1}-k_{2}$ plane.}
\label{figS6}
\end{figure*}

\begin{figure*}[htbp]
\begin{center}
\includegraphics[width=6.4in, clip=true]{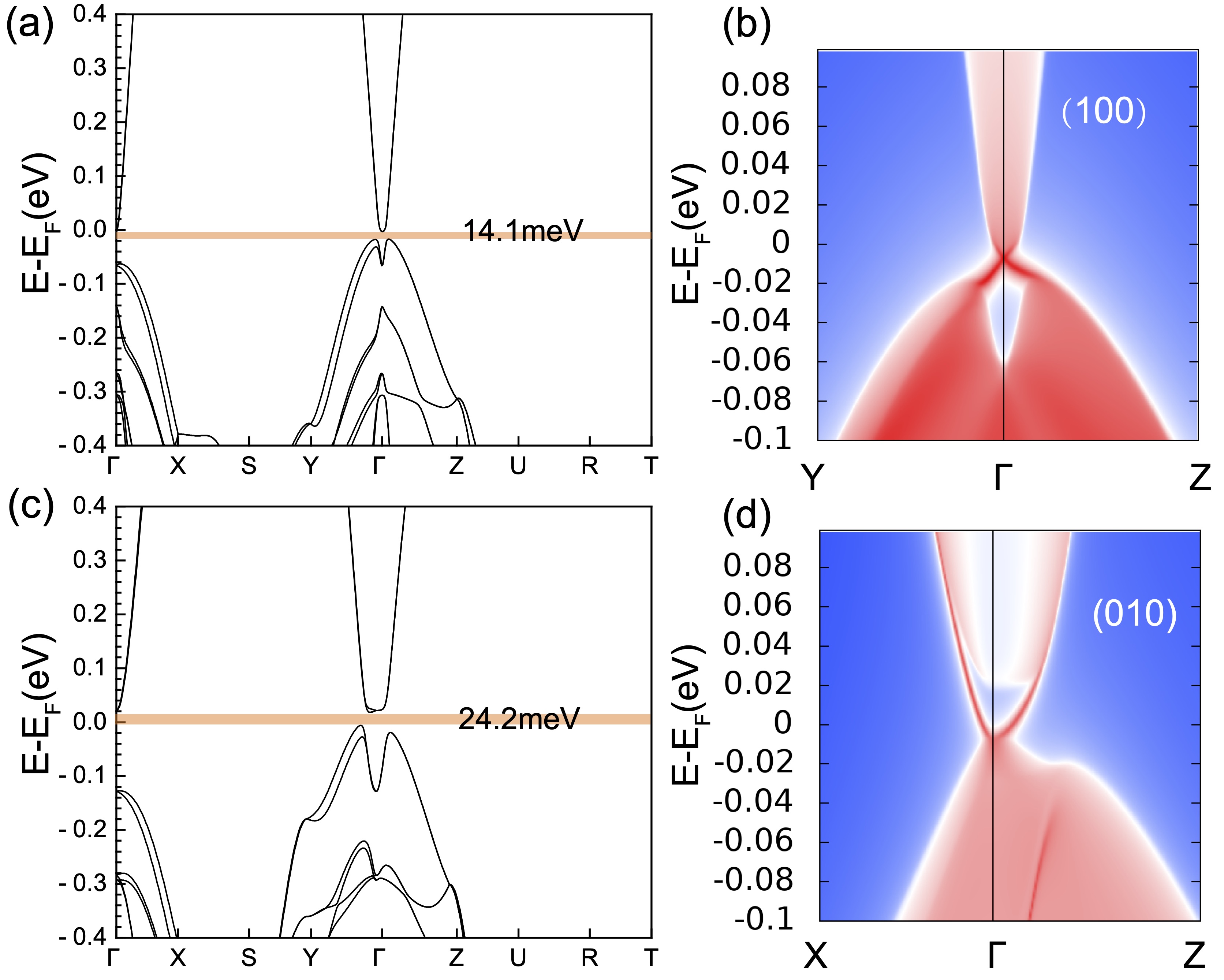}
\end{center}
\caption{Electron structures with SOC and surface states for Ag$_{2}$HgSe$_{4}$Sn. (a),(b) Resuls after adding Hubbard U correction (U$_{Ag}$=2~eV, U$_{Hg}$=1~eV). (c),(d) Results after full lattice constants relaxation.}
\label{figS7}
\end{figure*}

\begin{figure*}[htbp]
\begin{center}
\includegraphics[width=5.6in, clip=true]{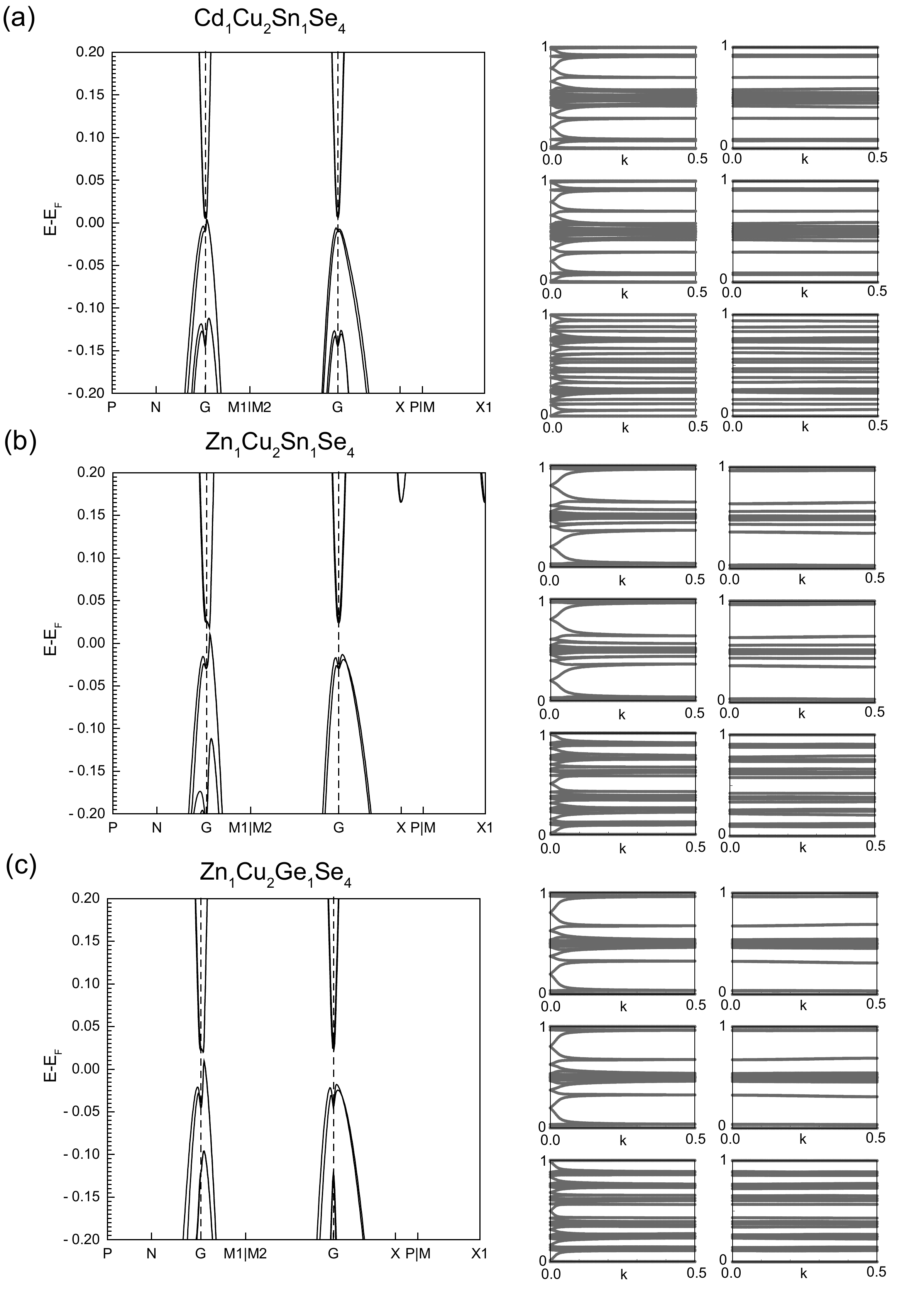}
\end{center}
\caption{Band structures and Wilson loops for (a)Cd$_{1}$Cu$_{2}$Sn$_{1}$Se$_{4}$; (b)Zn$_{1}$Cu$_{2}$Sn$_{1}$Se$_{4}$; (c)Zn$_{1}$Cu$_{2}$Ge$_{1}$Se$_{4}$}
\label{figS8}
\end{figure*}

\begin{figure*}[htbp]
\begin{center}
\includegraphics[width=5.6in, clip=true]{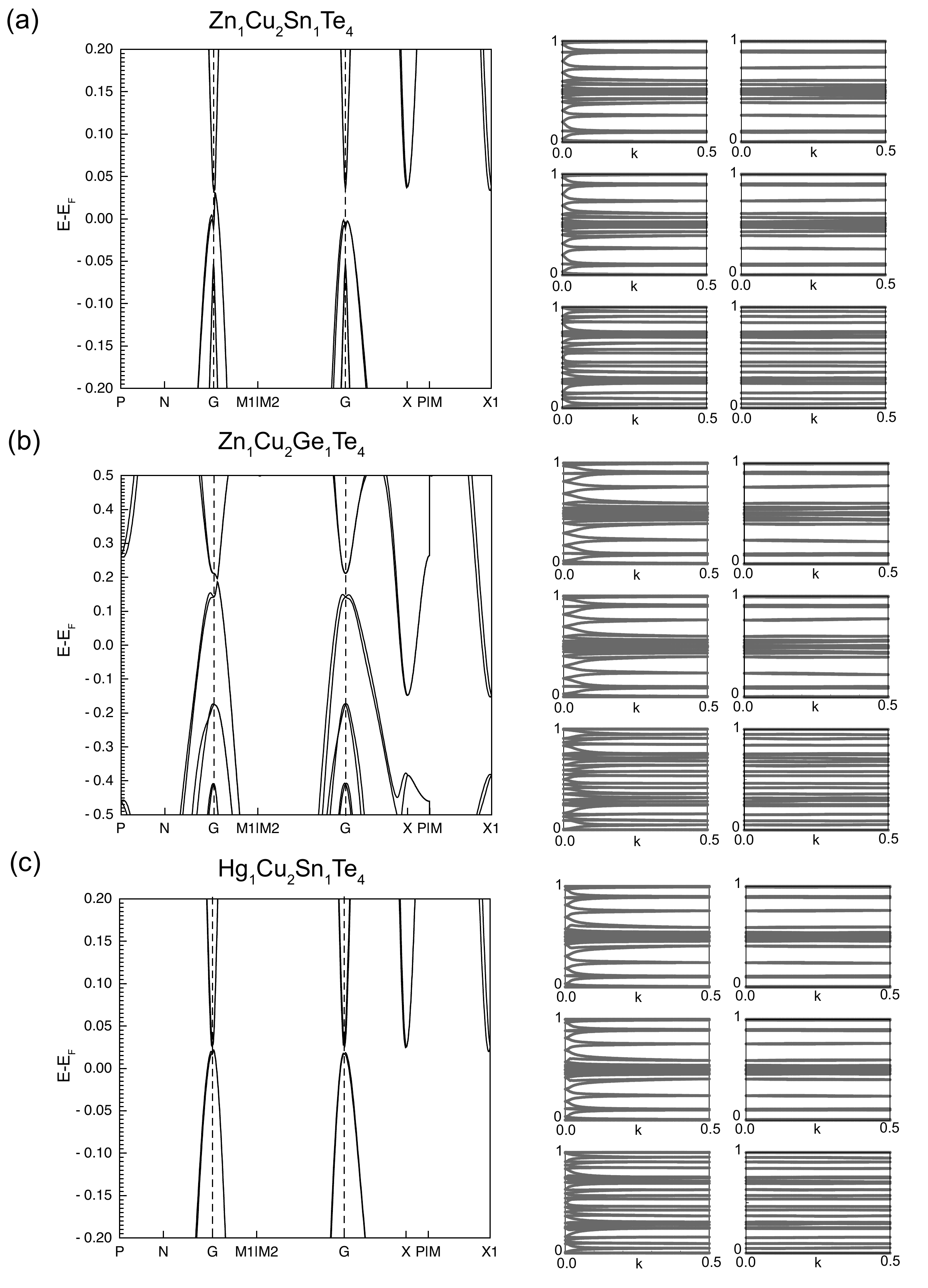}
\end{center}
\caption{Band structures and Wilson loops for (a)Zn$_{1}$Cu$_{2}$Sn$_{1}$Te$_{4}$; (b)Zn$_{1}$Cu$_{2}$Ge$_{1}$Te$_{4}$; (c)Hg$_{1}$Cu$_{2}$Sn$_{1}$Te$_{4}$}
\label{figS9}
\end{figure*}

\begin{figure*}[htbp]
\begin{center}
\includegraphics[width=5.6in, clip=true]{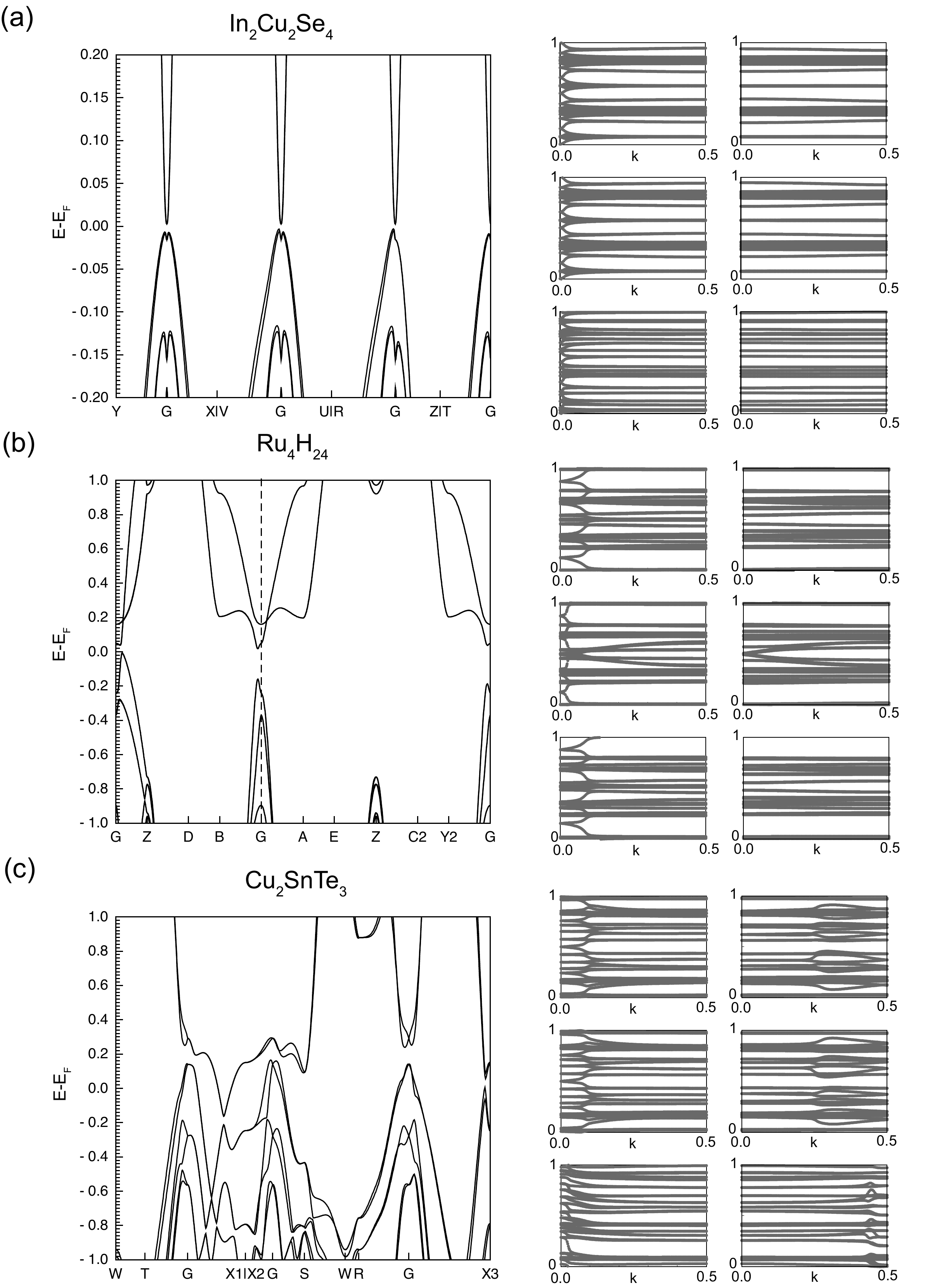}
\end{center}
\caption{Band structures and Wilson loops for (a)In$_{2}$Cu$_{2}$Se$_{4}$; (b)Ru$_{4}$H$_{24}$; (c)Cu$_{2}$SnTe$_{3}$}
\label{figS10}
\end{figure*}

\begin{figure*}[htbp]
\begin{center}
\includegraphics[width=5.6in, clip=true]{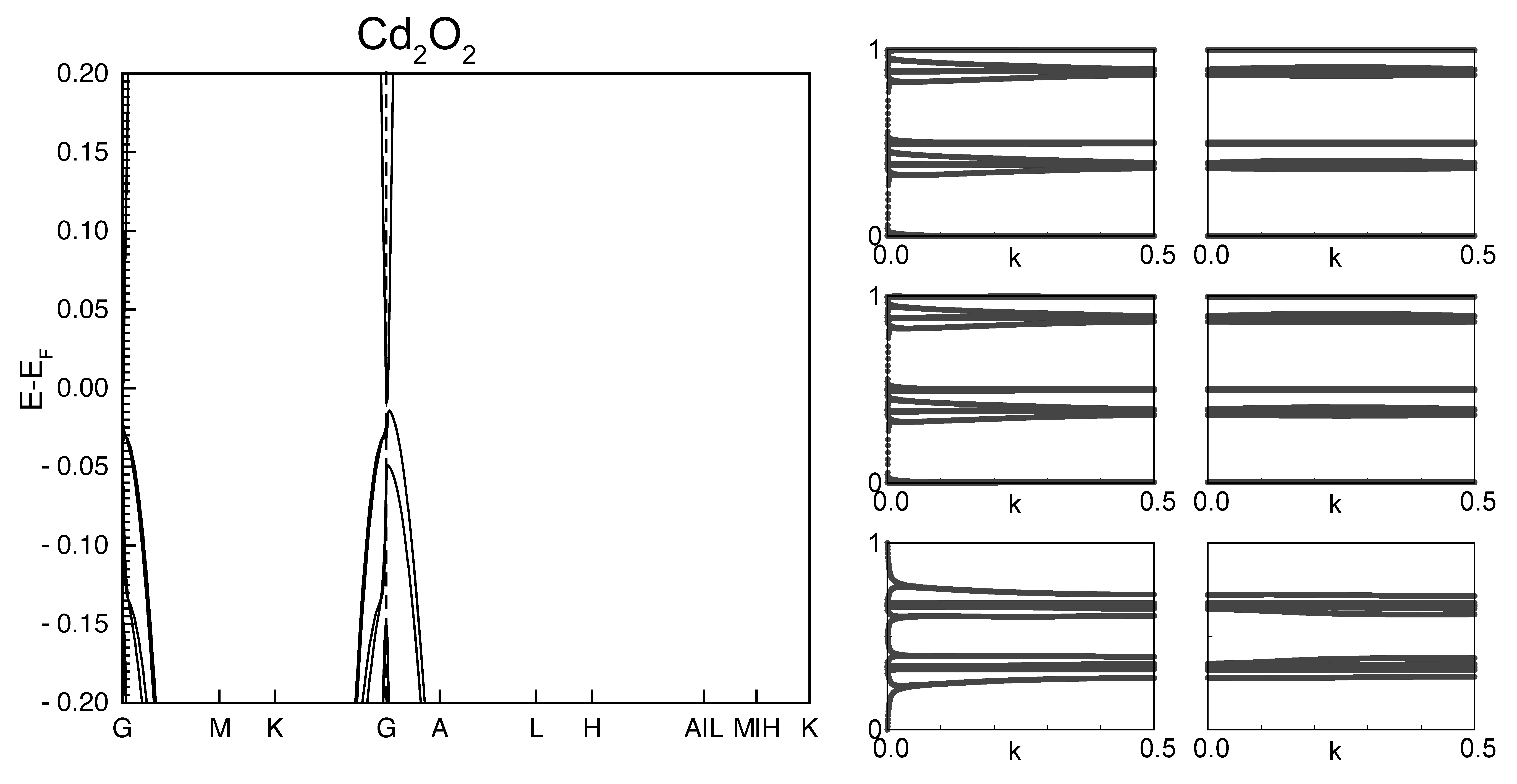}
\end{center}
\caption{Band structure and Wilson loop for Cd$_{2}$O$_{2}$}
\label{figS11}
\end{figure*}